\documentclass{article}

\usepackage[T1]{fontenc}
\usepackage[utf8]{inputenc}
\usepackage{lmodern}

\usepackage{amsmath}
\usepackage{amssymb}
\usepackage{mathrsfs}
\usepackage{bm}
\usepackage{physics}

\usepackage{graphicx}
\graphicspath{{figs/}}
\usepackage{array}
\usepackage{tabularx}
\usepackage{multirow}
\usepackage{subfigure}

\usepackage[english]{babel}
\usepackage{microtype}
\usepackage{verbatim}
\usepackage{blindtext}
\usepackage[normalem]{ulem}
\usepackage[dvipsnames]{xcolor}

\usepackage[numbers]{natbib}
\usepackage[colorlinks=true,
            linkcolor=black,
            citecolor=black,
            urlcolor=black]{hyperref}

\usepackage{cleveref}
\usepackage[all]{hypcap}

\usepackage{authblk}
\usepackage{orcidlink}

\title{Bounce, Turnaround, and the Anisotropy Problem in Cyclic Cosmology on a Brane with a Timelike Extra Dimension}

\author[1]{Rikpratik Sengupta\thanks{Email: rikpratik.sengupta@gmail.com}}
\author[2]{Arkajit Aich\thanks{Email: arkajit.aich.kolkata@gmail.com}}
\author[1]{Kaushik Bhattacharya\thanks{Email: kaushikb@iitk.ac.in}}

\affil[1]{Department of Physics, Indian Institute of Technology Kanpur, Kanpur 208016, Uttar Pradesh, India}
\affil[2]{Prayoga Institute of Education Research, Kanakpura Road, Bengaluru 562112, Karnataka, India}

\begin{document}
\maketitle

\begin{abstract}
We study cosmological bounces, turnarounds, and cyclic evolution on an
anisotropic Bianchi-I brane embedded in a five-dimensional bulk with a
\emph{timelike} extra dimension, within the Shtanov--Sahni braneworld
framework. Restricting to the flat, dark-radiation-free, effective-$\Lambda$-free
branch of the general anisotropic brane Friedmann equation, we drive the
dynamics with a single canonical scalar field obeying the uniform-rate
condition $\dot\phi=-\lambda=\mathrm{const}$, with shear anisotropy encoded
through a geometric term $\Omega_\sigma(a)\propto a^{-6}$. We derive a
general turning-point classification valid for any fluid obeying the null
energy condition: turnarounds at negative energy density occur
unconditionally, while bounces at $\rho>\rho_c$ occur only when the negative
high-energy brane correction dominates the decelerating shear term.
Specializing to the uniform-rate scalar, we obtain closed-form bounce and
turnaround conditions, the leading-order excess of the bounce density above
critical, and a matching condition for a finite cyclic branch connecting a
bounce at $N_B$ to a turnaround at $N_T$. We identify post-bounce
superinflationary and post-shear-dilution ordinary-inflationary regimes,
obtain a closed-form expression for the duration of the superinflationary
stage, $\Delta N_{\rm sinf}\simeq\rho_c/6\lambda^2=|\sigma|/3\lambda^2$, which
scales linearly with the brane tension, and show that the numbers of $e$-folds
separating the bounce, the pivot and the end of inflation are fixed
analytically by the tilt and the pivot density fraction alone,
compute the single-field curvature power spectrum, and derive parameter
relations fixing $H_*$, $\lambda$, $\rho_c$, and the shear amplitude
$\Sigma_g^2$ in terms of the observed amplitude $A_s$ and tilt $n_{s*}$. An
explicit CMB-normalized parameter point shows that sustaining a long,
weak-shear cyclic phase compatible with observations requires the shear
amplitude suppressed by $10^{2}$--$10^{3}$ orders of magnitude below
$H_*^2$, depending sensitively on the pivot density fraction
$x_*=\rho_{\phi*}/\rho_c$. We discuss the physical origin of this
anisotropy problem, its parametric dependence, and the status of the
periodicity condition required for a genuinely cyclic $V(\phi)$. In the
isotropic limit the construction reduces to an exactly solvable model whose
reconstructed periodic potential coincides, by the known duality between the
effective loop-quantum-cosmology and quadratic-brane Friedmann equations, with
the uniform-rate cyclic solution of loop quantum cosmology; the new content of
the present work lies entirely in the anisotropic sector.
\end{abstract}

\section{Introduction}

The presence of an initial spacetime singularity is one of the most persistent
conceptual shortcomings of the standard Big Bang framework and its modern
$\Lambda$CDM incarnation, even when supplemented by an early phase of cosmic
inflation. Under very general assumptions, namely the validity of classical
general relativity, reasonable energy conditions, and global hyperbolicity—the
Hawking–Penrose singularity theorems guarantee the existence of past-incomplete
geodesics, signaling a breakdown of the classical description of spacetime
\cite{HawkingEllis}. Inflation successfully addresses the horizon, flatness,
and monopole problems, and provides a compelling mechanism for generating
nearly scale-invariant primordial perturbations, yet it does not resolve the
initial singularity itself. In fact, inflationary spacetimes are generically
past-incomplete, as rigorously demonstrated by the Borde–Guth–Vilenkin theorem,
which shows that any universe undergoing sustained expansion must possess a
past boundary \cite{BGV}. However, it was recently shown that that inflationary spacetimes may remain geodesically complete in the infinite past even with sustained expansion, once the definition of average expansion and underlying topological assumptions are treated consistently \cite{Easson}. As a result, the inflationary paradigm still requires
an initial condition or a pre-inflationary phase that lies beyond its own
domain of validity.

Motivated by this limitation, a wide range of non-singular cosmological models
have been proposed over the decades, beginning with the steady-state theory of
Bondi, Gold, and Hoyle \cite{BondiGold,Hoyle}, and its later refinement in the
form of quasi–steady-state cosmology \cite{QSSC}. While these early models were
ultimately ruled out by observational evidence such as the cosmic microwave
background, they introduced the radical idea that cosmic evolution need not
originate from a singular beginning. More recent developments have led to
dynamical non-singular scenarios, including emergent universe models in which
the cosmos asymptotically approaches a static or quasi-static state in the
remote past before entering a phase of expansion \cite{EllisEmergent}, as well
as bouncing cosmologies where a prior contracting phase transitions smoothly
into the current expanding era \cite{NovelloReview,BrandenbergerReview}. In
cyclic models, this process is repeated indefinitely, replacing the notion of a
unique cosmic origin with an eternal sequence of expansions and contractions
\cite{SteinhardtTurok}.

Importantly, non-singular cosmologies can be realized both as extensions of the
inflationary framework and as genuine alternatives to it. In the former case,
a non-singular bounce or emergent phase precedes inflation, thereby providing
well-defined initial conditions for the inflationary epoch while avoiding a
past singularity \cite{BattefeldPeter}. In the latter approach, mechanisms such
as matter-dominated contraction, ekpyrotic phases, or modified gravity effects
generate the observed large-scale structure without invoking inflation at all
\cite{KhouryEkpyrotic,CaiBounce}. These scenarios often rely on violations of
classical energy conditions, higher-derivative corrections, quantum-gravity
effects, or extra-dimensional dynamics, highlighting the deep interplay between
early-universe cosmology and fundamental physics. Together, they illustrate
that resolving the initial singularity problem remains a central driver for
exploring physics beyond the standard cosmological model.

The idea that spacetime may possess more than four dimensions has long been a
source of theoretical motivation, originally emerging from attempts to unify
gravity with the other fundamental interactions. In modern high–energy physics,
extra dimensions arise naturally in string theory and related frameworks, where
they provide a consistent quantum description of gravity while offering new
mechanisms to address long–standing puzzles such as the hierarchy problem \cite{Pol}. From a cosmological perspective, extra dimensions open qualitatively new
possibilities for the early and late Universe, including modified expansion
laws, novel gravitational degrees of freedom, and dynamical mechanisms capable
of resolving classical singularities \cite{Gasperini}. Braneworld scenarios, in which our
observable Universe is confined to a lower–dimensional hypersurface embedded
in a higher–dimensional bulk, provide a particularly compelling realization of
these ideas, allowing higher–dimensional effects to manifest themselves in a
controlled and phenomenologically testable manner.

The earliest phenomenologically viable braneworld model was proposed by
Arkani-Hamed, Dimopoulos, and Dvali (ADD), who introduced large, compact
spacelike extra dimensions to dilute the fundamental Planck scale and thereby
address the gauge hierarchy problem \cite{ADD}. This was followed by the
Randall–Sundrum constructions, which demonstrated that strongly warped
geometries can localize gravity on a four–dimensional brane even when the extra dimension is non-compact \cite{Raman1,Raman2}. In the RS-I model, two branes of opposite tension bound a compact extra dimension, while RS-II retains a single positive-tension brane embedded in an infinite bulk with a negative cosmological constant. These scenarios lead to characteristic high–energy corrections to the Friedmann equations and have been extensively studied in early-universe cosmology, black hole physics, and gravitational phenomenology. In a series of recent works, one of the authors has investigated non-singular cosmological solutions as well as inflationary scenarios within braneworld frameworks \cite{RP1, RP2, RP3}. This single–brane setup with a spacelike extra dimension has been extensively
employed in a wide range of cosmological
studies~\cite{Binetruy,Maeda,Langlois,Chen,Kiritsis,Campos,Maartens}
as well as in diverse astrophysical
applications~\cite{Wiseman2,Germani,Deruelle,Wiseman,Visser,Creek,Pal,Bruni,Govender,Sengupta5}.

Subsequent developments extended the braneworld paradigm to incorporate
infrared modifications of gravity. The Dvali–Gabadadze–Porrati (DGP) model
introduced an induced Einstein–Hilbert term on the brane, leading to a crossover between four- and five-dimensional gravitational dynamics at large distances \cite{DGP}. This construction has important implications for late-time cosmology, including the emergence of self-accelerating solutions in the
absence of a cosmological constant, albeit at the cost of ghost instabilities.
A closely related, yet conceptually distinct, class of models was proposed by
Sahni and Shtanov, involving a spacelike extra dimension together with a
non-vanishing induced curvature term on the brane ($m\neq 0$) and a finite brane tension ($\sigma\neq 0$), which modifies the effective cosmological evolution \cite{SahniShtanov}. In this framework, both the strong and null energy conditions are violated at the level of the effective brane dynamics, naturally leading to late-time cosmic acceleration accompanied by an effective
phantom-like equation of state ($\omega_{\mathrm{eff}}<-1$). Importantly, this
behavior does not require the introduction of any fundamental phantom matter
component; as a result, the usual pathologies associated with ghosts and
instabilities in relativistic phantom models are avoided, and the accelerated
expansion arises purely from geometric modifications induced by the spacelike
extra dimension and the curvature term on the brane.

Of special relevance for early-universe cosmology is the Shtanov–Sahni scenario with a timelike extra dimension, in which the effective Friedmann equations on the brane acquire negative quadratic energy-density corrections
\cite{Sahni4}. These corrections generically lead to a cosmological
bounce at high energies, providing a natural and robust mechanism for avoiding
the initial Big Bang singularity without invoking exotic matter sources. When the extra dimension is spacelike, the bulk geometry possesses a Lorentzian
signature; in contrast, for a timelike extra dimension the bulk metric acquires
the signature $(-,-,+,+,+)$. In the latter case, the braneworld construction
necessarily involves a negative brane tension together with a positive bulk
cosmological constant, in sharp contrast to the Randall–Sundrum II scenario. A major challenge in models with timelike extra dimensions is the appearance of
tachyonic Kaluza–Klein graviton modes, which can potentially destabilize the
theory. This problem was resolved implementing a solitonic
three–brane configuration constructed within five–dimensional
Einstein–Hilbert–Gauss–Bonnet gravity formulated on a spacetime with a
non-Lorentzian signature containing an additional timelike direction \cite{S10}. In this
setup, the brane is represented as a $\delta$-function source that fully traps
the gravitational field, thereby eliminating all propagating degrees of freedom
in the bulk. As a consequence, gravitational dynamics on the brane reduce
exactly to four–dimensional Einstein gravity. Remarkably, the presence of a
timelike extra dimension does not introduce tachyonic excitations or
negative-norm states, ensuring the consistency and stability of the model. Timelike extra dimensions have also been investigated in other settings, notably in the construction of non-singular black hole and ``black universe'' solutions \cite{Bronnikov2006}, as well as exotic compact objects like wormholes \cite{RP4} and gravastars \cite{RP6}. A recent work shows that homogeneous dust collapse in the timelike Shtanov–Sahni braneworld avoids singularity formation due to finite brane tension keeping the curvature bounded \cite{RP5}. These studies highlight the Shtanov–Sahni construction as a particularly well–suited arena for exploring non-singular cosmologies, as it combines a clear geometric origin for the bounce with a controlled effective four–dimensional description that smoothly connects to general relativity at low energies.

In this paper we take the Shtanov--Sahni braneworld construction with a
timelike extra dimension already established as a natural,
matter-content-independent route to a non-singular bounce and ask
whether it can support a genuinely \emph{cyclic} history of the universe
once brane anisotropy, rather than isotropic matter alone, is allowed to
participate in the dynamics. Concretely, we generalize the effective
brane Friedmann equation to an anisotropic Bianchi-I geometry, sourced by
a single canonical scalar field evolving under the uniform-rate condition
$\dot\phi=\mathrm{const}$, and treat the shear as an additional,
purely geometric energy density diluting as $a^{-6}$. We derive, in
closed form and for a scalar field obeying the null energy condition,
the conditions under which the resulting cosmology bounces at high
density and turns around at low (formally negative) density, identify
the inflationary phases that arise naturally on the ensuing expansion
branch, and confront the model with CMB normalization to fix its free
parameters explicitly. The physical motivation for this program is
twofold. First, since Bianchi-I anisotropy is generically expected to
dominate the dynamics as $a\to0$ in any bouncing or cyclic scenario, it
is essential to establish, rather than assume away, whether the
negative high-energy correction responsible for the singularity-free
bounce in the timelike-extra-dimension braneworld can survive, and
continue to dominate, in the presence of shear; a bounce that exists only
in the idealized isotropic limit would not constitute a robust resolution
of the initial-singularity problem. Second, by demanding that the same
solution reproduce the observed amplitude and tilt of the primordial
curvature spectrum while also closing into a periodic $V(\phi)$ over
many hundreds of e-folds, we can quantify, rather than merely gesture at,
the well-known anisotropy problem of cyclic cosmologies within this
specific higher-dimensional framework, and thereby identify precisely
what additional physics, if any, such a model requires to be
observationally viable.

\section{Bounce, Turnaround and Cyclic Inflation on the Anisotropic (1+3)-Brane}
\label{sec:bounce-turnaround-cyclic}

In this section we study the anisotropic cosmological dynamics, coming out from Bianchi-I type metric on the (1+3)-brane. We primarily show that due to the effect of the (2+3)-dimensional theory the (1+3)-brane can accommodate a cyclic universe where each cycle can itself accommodate a phase of normal inflation. The main aim of this paper is to show that anisotropy can be properly reduced in the expanding phase  and one can have an normal inflationary  era in each expanding phase of a cycle. We do not intend to show how standard radiation domination appears after inflation as it requires more physics input. We have seen that this model can be enlarged and we can later try to address the cosmological phenomenology in more detail. To keep the model simple we have only considered one canonical scalar field on the brane as the sole matter component of the universe. 

\subsection{Matter sector and the effective Friedmann equation}
\label{subsec:setup}

The matter content is a single canonical scalar field $\phi(t)$ confined to the brane, with
\begin{equation}
\rho_\phi = \frac{1}{2}\dot\phi^2 + V(\phi), \qquad
p_\phi = \frac{1}{2}\dot\phi^2 - V(\phi),
\label{eq:rho-p-def}
\end{equation}
so that the null energy condition (NEC) holds identically:
\begin{equation}
\rho_\phi + p_\phi = \dot\phi^2 \ \geq\ 0 .
\label{eq:wec}
\end{equation}
We will later see that the scalar field potential may become sufficiently negative, in the expanding phase of a cycle, and $\rho_\phi <0$, for some brief time period, breaking the weak energy condition (WEC). At no time will the NEC be violated; the physical meaning of this brief negative-density excursion is examined in detail below Eq.~\eqref{eq:rhoT-approx}. The average scale factor of a Bianchi-I geometry, given by the line element:
$$ds^2=-dt^2 + \sum_{i=1}^3 a_i^2(t) (dx^i)^2\,,$$
is $a(t)=\left[a_1(t)a_2(t)a_3(t)\right]^{1/3}$, $H=\dot a/a$, where the anisotropic scale factors are given by $a_i$.
For vanishing anisotropic stress the shear scalar dilutes as $a^{-6}$; we encode this by the
purely geometric quantity
\begin{equation}
\Omega_\sigma(a) \ \equiv\ \frac{\Sigma_g^2}{a^6}, \qquad \Sigma_g^2 \ge 0 ,
\label{eq:omega-sigma-def}
\end{equation}
where $\Sigma_g$ is a constant.

The starting point is the general anisotropic Bianchi-I brane equation of Shtanov and Sahni, as obtained in \cite{Sahni4}:
\begin{equation}
H^2 
= \frac{\Lambda_{\rm eff}}{3} + \frac{8\pi G_N}{3}\,\rho
+ \epsilon\,\frac{\rho^2}{M^6} + \frac{C}{a^4} + \frac{\Sigma_g^2}{a^6},
\label{eq:SS-general}
\end{equation}
valid under the closure assumption that the projected nonlocal (Weyl) energy density on the
brane is negligible, $\epsilon=+1$ for a spacelike and $\epsilon=-1$ for a timelike extra
dimension, $C$ the dark-radiation integration constant, and $\Sigma_g^2\ge 0$ the shear
amplitude.

It should be emphasised explicitly that Eq.~\eqref{eq:SS-general} is not the most
general brane Friedmann equation of the Shtanov--Sahni class, but its
$m=0$ specialization. In the general Sahni--Shtanov construction
\cite{SahniShtanov} the brane carries, in addition to the tension $\sigma$, an
\emph{induced} four-dimensional Einstein--Hilbert term with coefficient $m^2$,
so that the action reads
$S=\epsilon M^3\big[\int_{\rm bulk}(\mathcal R-2\Lambda_b)-2\int_{\rm brane}K\big]
+\int_{\rm brane}(m^2R-2\sigma)+\int_{\rm brane}L$
and the brane Friedmann equation takes the ``two-branch'' square-root form
\begin{equation}
H^2 + \frac{\kappa}{a^2}
= \frac{8\pi}{3m^2}\big(\rho+\sigma\big) + \frac{2}{\ell^2}
\left[1 \pm \sqrt{\,1+\ell^2\left(\frac{8\pi}{3m^2}\big(\rho+\sigma\big)
- \frac{\Lambda_b}{6} - \frac{C}{a^4}\right)}\,\right],
\qquad \ell \equiv \frac{2m^2}{M^3},
\label{eq:SS-full}
\end{equation}
with $M$ the five-dimensional Planck mass and $\Lambda_b$ the bulk
cosmological constant. Setting $m=0$ (equivalently $\ell\to0$) removes the
induced-curvature term altogether, so that gravity on the brane is generated
purely by the five-dimensional bulk dynamics through the junction conditions.
In that limit the lower branch of Eq.~\eqref{eq:SS-full}, expanded to second
order in $\ell^2$, has its leading linear piece cancelled identically and
leaves
$H^2 = \Lambda_b/6 + C/a^4 + (64\pi^2/9M^6)(\rho+\sigma)^2$, i.e.\ precisely
the Randall--Sundrum-type quadratic form quoted in
Eq.~\eqref{eq:SS-general}, with $\Lambda_{\rm eff}$ and $G_N$ built out of
$\Lambda_b$ and $\sigma$ and with the sign $\epsilon$ of the $\rho^2$ term set
by the signature of the extra dimension \cite{Sahni4}. Since $G_N\propto\sigma$
in this identification, a positive effective Newton constant in the timelike
case is what forces $\sigma<0$, as used in Eq.~\eqref{eq:rhoc-def} below.
Throughout this paper we work exclusively in
this $m=0$, $\sigma\neq0$, $\epsilon=-1$ branch: there is no induced
Einstein--Hilbert term on the brane, no DGP-type infrared crossover, and
consequently no self-accelerating branch. All statements below --- in
particular the bounce condition and the turning-point classification --- refer
to this branch only, and would have to be re-derived for $m\neq0$, where the
high-energy behaviour of $H^2(\rho)$ is qualitatively different.

We specialize to the flat, timelike, dark-radiation-free, effective-$\Lambda$-free
branch relevant to uniform-rate cyclic inflation,
\begin{equation}
\epsilon = -1, \qquad C = 0, \qquad \Lambda_{\rm eff} = 0 .
\label{eq:minimal-choices}
\end{equation}
Defining the critical density
\begin{equation}
\rho_c \ \equiv\ \frac{8\pi G_N M^6}{3} = 2|\sigma|\,,
\label{eq:rhoc-def}
\end{equation}
($\sigma$ the brane tension, required negative for a positive effective Newton constant) and
setting $8\pi G_N = M_P^2 = 1$ throughout, Eq.~\eqref{eq:SS-general} reduces to:
\begin{equation}
\ H^2 = \frac{\rho_\phi}{3}\left(1 - \frac{\rho_\phi}{\rho_c}\right) + \Omega_\sigma(a)\,. 
\label{eq:master-friedmann}
\end{equation}
This is the effective average Friedmann equation for the geometric-anisotropy construction;
Eq.~\eqref{eq:minimal-choices}--\eqref{eq:master-friedmann} makes explicit that it is the
$C=0$, $\Lambda_{\rm eff}=0$, specialization of the general Shtanov--Sahni
Bianchi-I equation, closed under the vanishing-Weyl-term assumption. Any refinement that
restores $C$, $\Lambda_{\rm eff}$, or the nonlocal Weyl contribution would modify
Eq.~\eqref{eq:master-friedmann} and everything that follows.

Imposing a condition on the scalar field's evolution effectively fixes a corresponding form of the potential $V(\phi)$. In this work, instead of assuming a form of $V(\phi)$ to begin with, we will assume  uniform rate scalar evolution or a variant of it.
\subsection{Uniform-rate Scalar Evolution}

We impose the uniform-rate condition
\begin{equation}
\dot\phi = -\lambda, \qquad \lambda > 0, \qquad \ddot\phi = 0\,,
\label{eq:uniform-rate}
\end{equation}
where $\lambda$ is a real constant. Demanding a uniform-rate scalar evolution fixes the cyclicity of the emerging brane cosmology, but it also constrains the system. As we show below, this constraint limits the overall cosmological growth attainable in the expanding phase. We will generalize this condition to evade the problem. The Klein--Gordon equation $\ddot\phi + 3H\dot\phi + V_{,\phi} = 0$ then gives
\begin{equation}
V_{,\phi} = 3\lambda H .
\label{eq:KG-uniform}
\end{equation}
Energy conservation, $\dot\rho_\phi + 3H(\rho_\phi+p_\phi) = 0$, together with
Eq.~\eqref{eq:wec} evaluated on the uniform-rate trajectory ($\rho_\phi+p_\phi=\lambda^2$,
constant), gives $\dot\rho_\phi = -3\lambda^2 H$. In the $e$-fold variable $N\equiv\ln a$,
\begin{equation}
\frac{d\rho_\phi}{dN} = -3\lambda^2
\quad\Longrightarrow\quad
\rho_\phi(N) = A_0 - 3\lambda^2 N, \qquad
V(N) = A_0 - \frac{\lambda^2}{2} - 3\lambda^2 N,
\label{eq:rhophi-of-N}
\end{equation}
with $A_0$ an integration constant. This step uses only the conservation equation and
Eq.~\eqref{eq:wec}, not the specific form of the Friedmann equation, so it survives unchanged
from the isotropic uniform-rate brane model through to the anisotropic bounce/turnaround
setting of Eq.~\eqref{eq:master-friedmann}.


Before specializing to the uniform-rate scalar it is useful to establish the turning-point
structure of Eq.~\eqref{eq:master-friedmann} for a general perfect fluid satisfying
$\rho+p\ge 0$; the uniform-rate scalar is then simply the case $\rho+p=\lambda^2$.
A turning point of the mean scale factor occurs at $H_*=0$, i.e.\ at a root of
\begin{equation}
\frac{\rho_*}{3}\left(1-\frac{\rho_*}{\rho_c}\right) + \Omega_{\sigma *} = 0 .
\label{eq:turning-condition}
\end{equation}
Since $\Omega_{\sigma*}>0$ whenever $\Sigma_g\neq 0$, Eq.~\eqref{eq:turning-condition} forces
\begin{equation}
\rho_* < 0 \ \ \text{or}\ \ \rho_* > \rho_c .
\label{eq:two-branches}
\end{equation}
Conversely, for every $0<\rho<\rho_c$ both terms of Eq.~\eqref{eq:master-friedmann} are
strictly positive, so $H^2>0$ there: this interval is fully allowed dynamically and contains
no turning point. In particular the entire ordinary-inflationary regime discussed later in this
section, $0<\rho_\phi<\rho_c/2$, lies safely inside this
always-expanding-or-always-contracting region.

Differentiating Eq.~\eqref{eq:master-friedmann} and using $\dot\rho=-3H(\rho+p)$ together with
$\dot\Omega_\sigma = -6H\Omega_\sigma$ (immediate from $\Omega_\sigma\propto a^{-6}$) gives, for
$H\neq0$,
$$2H\dot H = -H(\rho+p)\Big(1-\frac{2\rho}{\rho_c}\Big) - 6H\Omega_\sigma \,,$$
and dividing by $2H$ before taking the limit to a regular turning point yields
\begin{equation}
\dot{H}_* = -\frac{1}{2}(\rho_*+p_*)\left(1-\frac{2\rho_*}{\rho_c}\right) - 3\,\Omega_{\sigma*} .
\label{eq:Hdot-general}
\end{equation}
This is the general classification formula, valid for any perfect fluid with $\rho+p\ge0$; it
can equivalently be obtained (and was cross-checked) by writing 
\begin{eqnarray}
H^2=F(N)\,, 
\end{eqnarray}
identically along the solution and using $\dot H=\tfrac12 F_{,N}$, extended by continuity to $F=0$. The two
independent derivations agree.

From the expressions of $H_*\,,\dot{H}_*$ one can calculate how a cosmological bounce and a turnaround happens. The energy density of the scalar sector at these two points will be denoted by $\rho_B$ and $\rho_T$, respectively.
If $\rho_T<0$ then $1-2\rho_T/\rho_c>1>0$, so together with $\rho_T+p_T\ge0$ the first term of
Eq.~\eqref{eq:Hdot-general} is non-positive, while $-3\Omega_{\sigma T}<0$ strictly. Hence
\begin{equation}
\dot H_T < 0 \qquad \text{(negative-density root: automatic turnaround)} .
\label{eq:HdotT-negative}
\end{equation}
The evolution is $H>0\to H=0\to H<0$: no tuning is required, and this holds for any fluid
obeying the null energy condition, not only the uniform-rate scalar. Here very near the turning point the scalar field kinetic term remains positive but the potential term becomes more negative compared to the kinetic term and the scalar field energy density becomes negative. This negative-density behavior is strictly local to the neighborhood of the turning point; throughout the inflationary regime itself, the scalar potential remains positive.

For a cosmological bounce we have to work in the high density branch of the theory. 
If $\rho_B>\rho_c$ then $1-2\rho_B/\rho_c<-1<0$, so the first term of
Eq.~\eqref{eq:Hdot-general} is now positive (again using $\rho_B+p_B\ge0$, with equality
excluded since $\dot\phi\neq0$), while the shear term $-3\Omega_{\sigma B}$ remains negative.
The root is a bounce, $\dot H_B>0$, iff
\begin{equation}
\frac{1}{2}(\rho_B+p_B)\left(\frac{2\rho_B}{\rho_c}-1\right) > 3\,\Omega_{\sigma B} .
\label{eq:bounce-condition-general}
\end{equation}
Unlike the turnaround, the bounce is therefore conditional: the negative high-energy brane
correction must dominate the decelerating shear term.
Specializing Eqs.~\eqref{eq:Hdot-general}--\eqref{eq:bounce-condition-general} to the
uniform-rate scalar, for which $\rho_\phi+p_\phi=\lambda^2$ identically, gives
\begin{equation}
\dot H_* = -\frac{\lambda^2}{2}\left(1-\frac{2\rho_{\phi*}}{\rho_c}\right) - 3\,\Omega_{\sigma*},
\label{eq:Hdot-uniform}
\end{equation}
with the bounce condition
\begin{equation}
\lambda^2\left(\frac{2\rho_{\phi B}}{\rho_c}-1\right) > 6\,\Omega_{\sigma B}
\quad\Longleftrightarrow\quad
\lambda^2\left(\frac{2\rho_{\phi B}}{\rho_c}-1\right) > 2\rho_{\phi B}\left(\frac{\rho_{\phi B}}{\rho_c}-1\right),
\label{eq:bounce-condition-uniform}
\end{equation}
the second form following from the Hamiltonian constraint
$\Omega_{\sigma B}=\tfrac{\rho_{\phi B}}{3}\big(\rho_{\phi B}/\rho_c-1\big)$ evaluated at the
bounce. This is the principal condition: shear cannot be arbitrarily large at the bounce, or
the anisotropic deceleration overwhelms the negative brane correction and no bounce occurs.

We now determine how far above $\rho_c$ the bounce occurs. Writing $x_B\equiv\rho_{\phi B}/\rho_c$ and $s_B\equiv\Omega_{\sigma B}$, the Hamiltonian
constraint at the bounce, $\tfrac{\rho_{\phi B}}{3}(1-\rho_{\phi B}/\rho_c)+\Omega_{\sigma B}=0$,
is the quadratic
\begin{equation}
x_B^2 - x_B - \frac{3 s_B}{\rho_c} = 0\,,
\label{eq:xB-exact}
\end{equation}
For weak shear, $s_B\ll\rho_c$, this gives the perturbative excess $x_B \simeq 1 + \frac{3 s_B}{\rho_c}$
which implies
\begin{eqnarray}
\rho_{\phi B} \ \simeq\ \rho_c + 3\,\Omega_{\sigma B}\,.
\label{eq:rhoB-approx}
\end{eqnarray}
Thus the anisotropic bounce need not occur at an arbitrarily large energy density of the scalar field: the excess above
the isotropic critical density $\rho_c$ is controlled by, and can be made perturbatively small
compared to, the isotropic critical density $\rho_c$ at the bounce.

The turnaround density follows from the Hamiltonian constraint. The same constraint at the turnaround, with $\rho_{\phi T}<0$ and $|\rho_{\phi T}|\ll\rho_c$ (in the limit, $1-\rho_{\phi T}/\rho_c\simeq1$), gives
\begin{equation}
\rho_{\phi T} \ \simeq\ -3\,\Omega_{\sigma T} .
\label{eq:rhoT-approx}
\end{equation}
A regular turnaround therefore corresponds to a small negative energy density, vanishing
in the limit $\Omega_{\sigma T}\to0$; it is never required to be large in magnitude.
Because the sign of $\rho_{\phi T}$ in Eq.~\eqref{eq:rhoT-approx} is central to the
consistency of the cyclic branch, it is worth stating explicitly what this negative
value means physically, and why it is not in conflict with the smallness of the
anisotropy. Three points should be distinguished.

\emph{(i) It is a negative potential energy, not negative matter.} The turnaround
occurs deep in the low-density regime, $|\rho_{\phi T}|\ll\rho_c$, where the quadratic
brane correction is utterly negligible and the Hamiltonian constraint
\eqref{eq:master-friedmann} reduces to $H^2\simeq\rho_\phi/3+\Omega_\sigma$. Since the
shear term is strictly positive, $H$ can vanish only if the scalar sector itself
contributes negatively, and Eq.~\eqref{eq:rhoT-approx} fixes by how much. Writing
$\rho_\phi=\tfrac12\lambda^2+V(\phi)$, the kinetic term is positive and \emph{constant}
along the uniform-rate trajectory, so the condition $\rho_{\phi T}<0$ is simply the
statement that the reconstructed potential has entered a negative region,
\begin{equation}
V(\phi_T) \;=\; \rho_{\phi T}-\frac{\lambda^2}{2}
\;\simeq\; -\,3\,\Omega_{\sigma T}-\frac{\lambda^2}{2} \;<\;0 .
\label{eq:VT-negative}
\end{equation}
There is no negative-mass, ghost or phantom matter anywhere in the construction: the
field is canonical, its kinetic energy never changes sign, and
$\rho_\phi+p_\phi=\lambda^2>0$ holds identically, so the NEC is respected at every
instant, as noted below Eq.~\eqref{eq:wec}. Only the WEC, $\rho_\phi\ge0$, fails, and
only within a narrow neighbourhood of the turning point. This is exactly the mechanism
by which a negative potential terminates expansion in ekpyrotic and cyclic scenarios
\cite{KhouryEkpyrotic,SteinhardtTurok,Ijjas}; here it is not postulated but follows
from the reconstruction, the isotropic closed form \eqref{eq:Vphi-isotropic} exhibiting
explicitly a potential whose minimum value is $-\lambda^2/2<0$.

\emph{(ii) The small anisotropy makes the effect smaller, not larger.} The referee's
concern, that a negative $\rho_{\phi T}$ is puzzling when the anisotropy is tending to
zero, is resolved by noting that Eq.~\eqref{eq:rhoT-approx} makes the two vanish
\emph{together}: $\rho_{\phi T}=-3\Omega_{\sigma T}\to0^-$ as $\Omega_{\sigma T}\to0$.
The required negative density is therefore not an independent assumption but is slaved
to the residual shear, and it is parametrically as small as the shear itself. For the
CMB-normalized benchmark of Sec.~\ref{subsec:strategy}, the turnaround shear is
$\Omega_{\sigma T}=\Omega_{\sigma B}e^{-6\Delta N_{\rm half}}\sim10^{-1346}$ in reduced
Planck units, so that $|\rho_{\phi T}|/\rho_c\sim10^{-1339}$: the ``negative energy
density'' at the turnaround is, numerically, indistinguishable from zero. In the strict
isotropic limit $\Sigma_g=0$ it is exactly zero, and the turnaround takes place at
$\rho_\phi=0$ precisely (Sec.~\ref{subsec:isotropic-limit}). Hence nothing discontinuous
or singular happens as the anisotropy is switched off; Eq.~\eqref{eq:rhoT-approx}
interpolates smoothly to the isotropic result.

\emph{(iii) It is local and transient.} The total effective energy density driving
the expansion, $\rho_{\rm tot}\equiv\rho_\phi+3\Omega_\sigma$, is by construction
\emph{zero} at the turnaround and positive on either side of it, so the geometry is
perfectly regular there: $H$ passes through zero with $\dot H_T<0$ fixed by
Eq.~\eqref{eq:HdotT-negative}, and $a(t)$ has a smooth maximum. The excursion
$\rho_\phi<0$ lasts only for the brief interval around $N_T$ during which
$|\rho_\phi|\lesssim3\Omega_\sigma$; away from that window, and in particular throughout
the entire observationally relevant inflationary stage, $\rho_\phi>0$ and $V>0$.
Physically, the turnaround is thus driven by the scalar rolling into a shallow negative
trough of its own reconstructed potential, which converts the expansion into contraction
and initiates the next half-cycle --- the low-density counterpart of the high-density
bounce driven by the negative $\rho^2$ brane correction.

\subsubsection{On a finite cyclic branch with normal inflation} 
\label{subsubsec:cyclic-branch}

\begin{figure}[t]
\centering
\includegraphics[width=\linewidth]{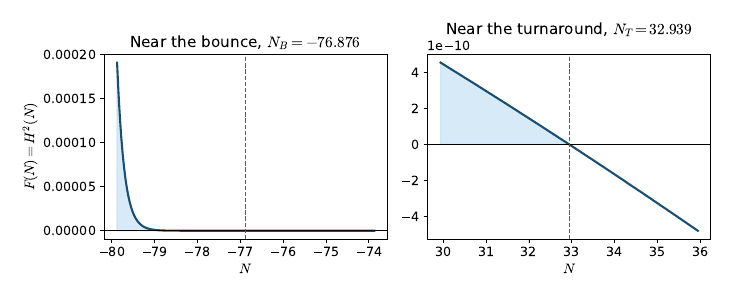}
\caption{$F(N)=H^2(N)$ in the immediate vicinity of the bounce (left) and the
turnaround (right), confirming $F>0$ on the finite interval between the two roots.
}
\label{fig:turning-points}
\end{figure}
If, in a cycle, the number of $e$-folds up to a bounce is $N_B$ and the number of $e$-folds up to the consequent turnaround is $N_T$ then a cyclic branch exists if there are two adjacent roots $N_B<N_T$ of
Eq.~\eqref{eq:turning-condition}. With $F(N)>0$ for $N_B<N_T$ interior (guaranteed by
the previous discussion), $F_{,N}(N_B)>0$ (bounce) and $F_{,N}(N_T)<0$ (turnaround).
Using $\rho_{\phi T}=\rho_{\phi B}-3\lambda^2\Delta N_{\rm half}$ where the half-cycle $e$-fold number is $\Delta N_{\rm half}\equiv N_T-N_B$, and
$\Omega_{\sigma T}=\Omega_{\sigma B}\,e^{-6\Delta N_{\rm half}}$, the two Hamiltonian constraints combine
into the matching condition
\begin{equation}
e^{-6\Delta N_{\rm half}} = \frac{-\rho_{\phi T}\big(1-\rho_{\phi T}/\rho_c\big)}{\rho_{\phi B}\big(\rho_{\phi B}/\rho_c-1\big)} .
\label{eq:matching-condition}
\end{equation}
A large half-cycle $\Delta N_{\rm half}$ requires the right-hand side to be exponentially small, which can
be arranged either by taking the turnaround energy density $\rho_{\phi T}\to0^-$
(Eq.~\eqref{eq:rhoT-approx}, i.e.\ weak shear at turnaround) or by pushing the bounce energy density
sufficiently far above $\rho_c$ (Eq.~\eqref{eq:rhoB-approx}), or a combination of both. Since
Eqs.~\eqref{eq:rhoB-approx}--\eqref{eq:rhoT-approx} give both densities directly in terms of the
shear amplitudes at the two turning points, Eq.~\eqref{eq:matching-condition} becomes, in the
weak-shear regime, an explicit relation between $\Omega_{\sigma B}$, $\Omega_{\sigma T}$,
$\rho_c$, and $\Delta N_{\rm half}$, rather than an implicit one.

\begin{figure}[!htbp]
\centering
\includegraphics[width=0.85\linewidth]{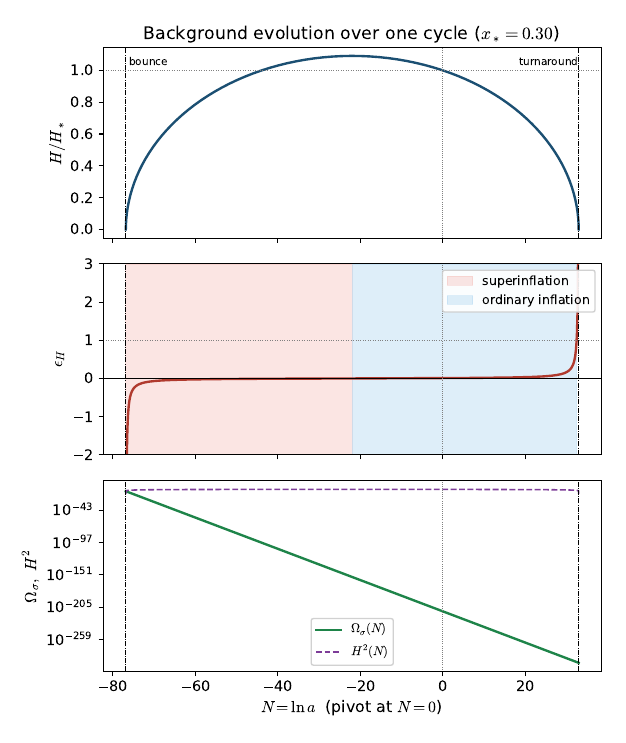}
\caption{Background evolution over one full cycle at $x_*=0.30$: normalized Hubble
rate $H/H_*$ (top), Hubble-flow parameter $\epsilon_H$ with the superinflationary
($\epsilon_H<0$) and ordinary-inflationary ($0<\epsilon_H<1$) regions shaded (middle),
and the shear density $\Omega_\sigma(N)$ compared to $H^2(N)$ (bottom). Vertical
dash-dotted lines mark the bounce $N_B$ and turnaround $N_T$; the dotted line marks
the CMB pivot $N=0$.}
\label{fig:cycle-dynamics}
\end{figure}

After the bounce, in the expansion branch, we have inflation. On the expansion branch, $H>0$, and we define the Hubble-flow parameter (the first slow-roll parameter):
\begin{equation}
\epsilon_H \equiv -\frac{\dot H}{H^2}
= \frac{\tfrac12\lambda^2\big(1-2\rho_\phi/\rho_c\big) + 3\Omega_\sigma}
{\tfrac13\rho_\phi\big(1-\rho_\phi/\rho_c\big) + \Omega_\sigma} .
\label{eq:epsilonH}
\end{equation}
Cold inflation happens when, $\ddot a/a>0$, and $0<\epsilon_H<1$. 
It is known that at the bounce point, $\dot H_B>0$ and by continuity there is a finite interval immediately
after the bounce with $\epsilon_H<0$ ($H>0$, $\dot H>0$). This period is accompanied by accelerated expansion of the system which we call superinflation and is different from the standard inflationary phase which follows. During this superinflation phase the shear energy gets diluted continuously and then later a phase of normal inflation begins.

The duration of this superinflationary stage can be given in closed form, and it is
useful to do so because it is this quantity, rather than the bounce itself, that
controls how much of the expanding branch is \emph{not} ordinary inflation. From
Eq.~\eqref{eq:Hdot-uniform}, $\epsilon_H<0$ holds as long as
\begin{equation}
\lambda^2\left(\frac{2\rho_\phi}{\rho_c}-1\right) \;>\; 6\,\Omega_\sigma(N) .
\label{eq:superinf-window}
\end{equation}
The bounce condition \eqref{eq:bounce-condition-uniform} is precisely the statement that
Eq.~\eqref{eq:superinf-window} is satisfied at $N=N_B$. Immediately afterwards the two
sides evolve in opposite directions and on very different timescales: the right-hand
side collapses exponentially, $\Omega_\sigma\propto e^{-6(N-N_B)}$, whereas the
left-hand side decreases only linearly, since $\rho_\phi=\rho_{\phi B}-3\lambda^2(N-N_B)$
by Eq.~\eqref{eq:rhophi-of-N}. Consequently the shear term drops out of
Eq.~\eqref{eq:superinf-window} within a few $e$-folds of the bounce, and the
superinflationary stage is terminated not by the shear but by the slow linear decrease
of $\rho_\phi$ through the value $\rho_c/2$, at which the brane correction ceases to
dominate. Setting $\rho_\phi=\rho_c/2$ in Eq.~\eqref{eq:rhophi-of-N} and using the
weak-shear bounce density \eqref{eq:rhoB-approx}, $\rho_{\phi B}\simeq\rho_c+3\Omega_{\sigma B}$,
gives the explicit duration
\begin{equation}
\Delta N_{\rm sinf} \;\equiv\; N_{\rm sinf}-N_B
\;=\; \frac{\rho_{\phi B}-\rho_c/2}{3\lambda^2}
\;\simeq\; \frac{\rho_c}{6\lambda^2}\left(1+\frac{6\,\Omega_{\sigma B}}{\rho_c}\right)
\;\simeq\; \frac{\rho_c}{6\lambda^2}\;=\;\frac{\Delta N_{\rm half}}{2},
\label{eq:superinf-duration}
\end{equation}
where $\Delta N_{\rm half}=\rho_c/3\lambda^2$ is the half-cycle length obtained below
in Sec.~\ref{subsec:strategy}. Thus the answer to ``how long, and controlled by what''
is unambiguous: to leading order in the weak-shear regime the superinflationary phase
occupies exactly the \emph{first half} of the expanding branch, namely the entire
high-density segment $\rho_c/2<\rho_\phi<\rho_{\phi B}$, and its length is set by the
single dimensionless combination $\rho_c/\lambda^2$. The shear enters only through the
positive fractional correction $6\Omega_{\sigma B}/\rho_c$, which for the CMB-normalized
parameter point of Sec.~\ref{subsec:strategy} is $\mathcal{O}(10^{-4})$ and hence
entirely negligible; the qualification that the shear could shorten this stage applies
only in the marginal regime $6\Omega_{\sigma B}\to\lambda^2$, where the bounce inequality
\eqref{eq:bounce-condition-uniform} is close to saturation and superinflation is indeed
squeezed to zero duration. For the benchmark of Eq.~\eqref{eq:worked-cmb} ($x_*=0.1$),
Eq.~\eqref{eq:superinf-duration} gives $\Delta N_{\rm sinf}\simeq256$, in agreement with
the numerically shaded region of Fig.~\ref{fig:cycle-dynamics}.

Equation~\eqref{eq:superinf-duration} also makes the dependence on the brane tension
completely transparent. Using $\rho_c=2|\sigma|$ from Eq.~\eqref{eq:rhoc-def},
\begin{equation}
\Delta N_{\rm sinf} \;\simeq\; \frac{|\sigma|}{3\lambda^2},
\qquad
\rho_{\phi B}\simeq\rho_c=2|\sigma|,
\qquad
\Omega_{\sigma B} < \frac{\lambda^2}{6},
\label{eq:sigma-scaling}
\end{equation}
so that at fixed scalar velocity $\lambda$ the superinflationary stage lengthens
\emph{linearly} with the magnitude of the brane tension, as does the full half-cycle
$\Delta N_{\rm half}=2|\sigma|/3\lambda^2$, while the bounce density is simply
$2|\sigma|$. It is worth noting the contrast with the upper bound on the shear at the
bounce, $\Omega_{\sigma B}<\lambda^2/6$, which is manifestly \emph{independent} of
$\sigma$: the brane tension controls how long the high-energy stage lasts and at what
density the bounce occurs, but not how much anisotropy the bounce can tolerate. A larger
$|\sigma|$ therefore does not relax the anisotropy problem quantified in
Sec.~\ref{subsec:strategy}; on the contrary, by lengthening $\Delta N_{\rm half}$ it
increases the $e^{6\Delta N_{\rm half}}$ re-amplification of shear across a cycle and
makes the required suppression of $\Sigma_g^2$ more severe. This inverse trade-off
between brane tension and anisotropy tolerance is a structural feature of the model.

$\dot H$ changes sign, keeping $H>0$, when
$\lambda^2(2\rho_\phi/\rho_c-1)=6\Omega_\sigma$ [the boundary of
Eq.~\eqref{eq:Hdot-uniform}]; from and after this point, we have $\dot H \le 0$, but acceleration can persist for
$0<\epsilon_H<1$. Once shear is diluted, $\Omega_\sigma\ll H^2$, Eq.~\eqref{eq:epsilonH}
reduces to the isotropic uniform-rate expression
\begin{equation}
\epsilon_H \ \simeq\ \frac{3\lambda^2\big(1-2\rho_\phi/\rho_c\big)}{2\rho_\phi\big(1-\rho_\phi/\rho_c\big)} ,
\qquad 0<\rho_\phi<\frac{\rho_c}{2},\qquad 0<\epsilon_H\ll1,
\label{eq:epsilonH-isotropic}
\end{equation}
which  sits entirely inside the always-allowed interval
$0<\rho_\phi<\rho_c$ and is therefore free of any turning-point subtlety. This is the phase in
which the single-clock, single-field power spectrum of Sec.~\ref{subsec:spectrum} applies.

Since $\Omega_\sigma\propto a^{-6}$, shear dilutes during the expansion branch but is
amplified by the same factor during the subsequent contraction. If the half-cycle spans
$\Delta N_{\rm half}$, e-folds,
the shear inherited at the next bounce must still satisfy
the bounce condition Eq.~\eqref{eq:bounce-condition-uniform} there, i.e.
\begin{equation}
\Omega_{\sigma T}\, e^{6 \Delta N_{\rm half}} \ <\ \frac{\lambda^2}{6}\left(\frac{2\rho_{\phi B}}{\rho_c}-1\right) .
\label{eq:cyclic-shear-bound}
\end{equation}
Using the weak-shear bounce relation Eq.~\eqref{eq:rhoB-approx}, the right-hand side reduces to
$\simeq \tfrac{\lambda^2}{6}$ at leading order (independent of $\Omega_{\sigma B}$ itself),
so Eq.~\eqref{eq:cyclic-shear-bound} is primarily a bound on how much shear the turnaround is
allowed to carry, given $\Delta N_{\rm half}$. This is the familiar anisotropy problem of bouncing and
cyclic cosmologies, expressed here in closed form in terms of $\lambda$, $\rho_c$, and the
turning-point shear amplitudes.


The scalar field is strictly monotonic, $\phi(t)=\phi_0-\lambda t$, while $\rho_\phi$ and $V(\phi)$ depend
only on $N$ through Eq.~\eqref{eq:rhophi-of-N}. The field-space distance covered between a
bounce and the following turnaround is
\begin{equation}
\Delta\phi_{B\to T} = \lambda\int_{N_B}^{N_T}\frac{dN}{\sqrt{F(N)}}, \qquad
\Delta\phi_{\rm cyc} = 2\lambda\int_{N_B}^{N_T}\frac{dN}{\sqrt{F(N)}} ,
\label{eq:phi-cyc}
\end{equation}
and since $\oint H\,dt=\Delta\ln a=0$ over one full cycle, Eq.~\eqref{eq:KG-uniform} gives
\begin{equation}
\Delta V_{\rm cyc} = \int V_{,\phi}\,\dot\phi\,dt = -3\lambda^2\oint H\,dt = 0 ,
\label{eq:DeltaVcyc}
\end{equation}
with $V_{,\phi}=0$ at both turning points ($H=0$ in Eq.~\eqref{eq:KG-uniform}).

Because $a(t)$ (hence $\rho_\phi$ and $V$ as functions of $N$) is periodic while $\phi(t)$ is
monotonic, the \emph{same} value of $\Delta N$ and therefore the same magnitude of $\Delta \rho_\phi$ and the same
magnitude of $\Delta V$ is revisited once on the expanding half-cycle and once on the contracting half-cycle, at
two distinct field values separated by $\Delta\phi_{\rm cyc}$. Equations~\eqref{eq:phi-cyc}--\eqref{eq:DeltaVcyc}
show that the net change of $V$ around one cycle vanishes and that $V$ has stationary points at
the turning points; they do not by themselves guarantee that $\phi\mapsto V(\phi)$ is a single
well-defined, smooth periodic function. This requires the additional consistency condition
\begin{equation}
V(\phi+\Delta\phi_{\rm cyc}) = V(\phi) \qquad \text{for all } \phi,
\label{eq:periodicity-condition}
\end{equation}
together with matching of $V_{,\phi}$ (equivalently, of $H$ via Eq.~\eqref{eq:KG-uniform}) at
the two field values corresponding to the same $N$ on the up- and down-crossings. We record
Eq.~\eqref{eq:periodicity-condition} as a condition to be verified explicitly once a specific
$\Sigma_g$, $\rho_c$, $\lambda$ parameter points are explicitly chosen. 

\subsection{Scalar power spectrum in the ordinary-inflationary regime}
\label{subsec:spectrum}

We now compute the approximate form of the standard cold-inflationary power spectrum in this framework.
In the regime $\Omega_\sigma\ll H^2$, perturbation
theory is effectively single-field and isotropic. With
$P_{\delta\phi}\simeq(H_*/2\pi)^2$ and $\dot\phi^2=\lambda^2$, the curvature spectrum on the sheer-diluted approximate FLRW system on the Bianchi-I brane is:
\begin{equation}
\mathcal P_{\mathcal R}(k) \ \simeq\ \frac{H^4}{4\pi^2\lambda^2}\bigg|_{k=aH}
\ =\ \frac{1}{4\pi^2\lambda^2}\left[\frac{\rho_\phi}{3}\Big(1-\frac{\rho_\phi}{\rho_c}\Big)+\Omega_\sigma\right]^2\bigg|_{k=aH},
\label{eq:PR}
\end{equation}
reducing, when shear is negligible at horizon exit, to
\begin{equation}
\mathcal P_{\mathcal R}(k) \ \simeq\ \frac{\rho_\phi^2\big(1-\rho_\phi/\rho_c\big)^2}{36\pi^2\lambda^2}\bigg|_{k=aH} \,.
\label{eq:PR-isotropic}
\end{equation}
Here $k$ is the wave number of the perturbation. Standard inflationary theory gives $n_s-1=d \ln \mathcal P_{\mathcal R}(k)/d \ln k$. At horizon crossing $k=aH$ and consequently $\ln k = N + \ln H$, hence $d \ln k/d N = 1 + (d\ln H/d N)=1-\epsilon_H$ where $\epsilon_H=-d\ln H/d N$. As $\mathcal P_{\mathcal R}(k) \propto H^4$ we have $d\ln\mathcal P_{\mathcal R}/dN=-4\epsilon_H$ and from the form of $d\ln k/dN$ at horizon
crossing, the tilt is given by:
$$n_s - 1 = \frac{-4\epsilon_H}{1-\epsilon_H}\,,$$
yielding
\begin{eqnarray}
\epsilon_{H*} = \frac{1-n_{s*}}{5-n_{s*}} \ \approx\ 8.7\times10^{-3}
\quad (n_{s*}\simeq0.965) ,
\label{eq:tilt}
\end{eqnarray}
up to corrections from residual anisotropy and non-standard initial states. Here we have used the value of $n_s$ obtained from Planck 2018 results \cite{Planck2018Inflation}. Differentiating
Eq.~\eqref{eq:tilt} gives the running,
\begin{equation}
\alpha_s \equiv \frac{dn_s}{d\ln k} = -\frac{4\,\epsilon_{H,N}}{(1-\epsilon_H)^3},
\qquad \epsilon_{H,N}\equiv\frac{d\epsilon_H}{dN},
\label{eq:running}
\end{equation}
small whenever $\epsilon_H$ varies slowly over the observable range, i.e.\ $|\alpha_s|\ll|n_s-1|$.

\begin{figure}[t]
\centering
\includegraphics[width=0.75\linewidth]{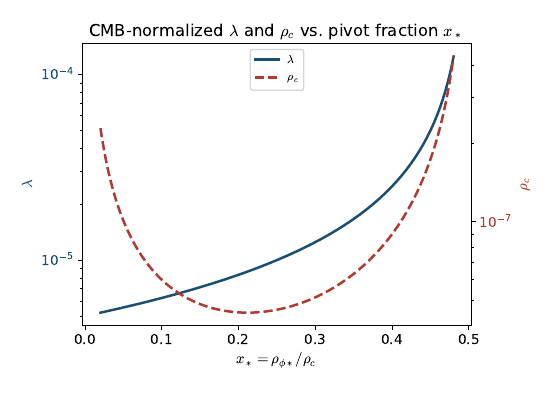}
\caption{CMB-normalized $\lambda$ and $\rho_c$ as functions of the pivot density
fraction $x_*=\rho_{\phi*}/\rho_c$, fixed by the observed amplitude $A_s$ and tilt
$n_{s*}$ via Eqs.~\eqref{eq:Hstar}--\eqref{eq:rhoc-fit}.}
\label{fig:lambda-rhoc}
\end{figure}

Writing the power spectrum in the way
$$\mathcal P_{\mathcal R}(k) = A_s \left(\frac{k}{k_*}\right)^{n_s -1 + \frac12 \alpha_s \ln(k/k_*)+\cdot\cdot\cdot}\,,$$
where $k_*=0.05\,{\rm Mpc}^{-1}$ is pivot scale used by Planck. Planck gives the amplitude of the power spectrum at the pivot point as $A_s\simeq2.1\times10^{-9}$. Let $x_*\equiv\rho_{\phi*}/\rho_c$ and $\delta_*\equiv\Omega_{\sigma*}/H_*^2\ll1$
(residual shear fraction at the pivot). The pivot Friedmann equation and
Eq.~\eqref{eq:epsilonH} give
\begin{equation}
H_*^2(1-\delta_*) = \frac{\rho_c\,x_*(1-x_*)}{3}, \qquad
\epsilon_{H*} = \frac{\lambda^2}{2H_*^2}(1-2x_*) + 3\delta_* .
\label{eq:pivot-eqs}
\end{equation}
The second equation above yields
\begin{eqnarray}
\lambda^2 = \frac{2H_*^2(\epsilon_{H*} - 3\delta_*)}{1-2x_*}\,,
\end{eqnarray}
as a consequence positivity of $\lambda^2$ requires $\delta_*<\epsilon_{H*}/3\approx2.9\times10^{-3}$. 
For a fully consistent approach we assume that the isotropic-looking spectrum in practice requires the much stronger condition $\delta_*\ll10^{-3}$. This condition is only a stronger version of the bound we obtained above.
Combining Eq.~\eqref{eq:pivot-eqs} with the amplitude $A_s\equiv\mathcal P_{\mathcal R}(k_*)=H_*^4/(4\pi^2\lambda^2)$
gives
\begin{align}
H_*^2 &= \frac{8\pi^2 A_s(\epsilon_{H*}-3\delta_*)}{1-2x_*}, \label{eq:Hstar} \\
\lambda^2 &= \frac{16\pi^2 A_s(\epsilon_{H*}-3\delta_*)^2}{(1-2x_*)^2}, \label{eq:lambda2} \\
\rho_c &= \frac{24\pi^2 A_s(1-\delta_*)(\epsilon_{H*}-3\delta_*)}{x_*(1-x_*)(1-2x_*)}, \label{eq:rhoc-fit}\\
\Omega_{\sigma*} &= \delta_* H_*^2 = \frac{8\pi^2 A_s\,\delta_*(\epsilon_{H*}-3\delta_*)}{1-2x_*}, \qquad
\Sigma_g^2 = a_*^6\,\Omega_{\sigma*}. \label{eq:sigma-fit}
\end{align}
The first of the above four equations shows that $0 < x_* < 1/2$.
With the fiducial Planck normalization $A_s\simeq2.1\times10^{-9}$ and
$\epsilon_{H*}\approx8.7\times10^{-3}$, Eqs.~\eqref{eq:Hstar}--\eqref{eq:sigma-fit} fix
$H_*$, $\lambda$ and $\rho_c$ 
once $x_*$ and $\delta_*$ are chosen.

The observable anisotropy parameter at the pivot is \(\delta_*=\Omega_{\sigma *}/H_*^2\), with \(\Omega_{\sigma *}=\Sigma_g^2/a_*^6\). Hence the CMB normalization constrains the physical shear contribution \(\Omega_{\sigma *}\),
rather than the integration constant \(\Sigma_g\) alone. If one fixes the scale-factor normalization by setting \(a_*=1\), then \(\Sigma_g^2=\Omega_{\sigma *}\); otherwise \(\Sigma_g^2=\Omega_{\sigma *}a_*^6\).

\subsubsection{Calibrating the theory with the CMB-normalized parameter point}
\label{subsec:strategy}

\begin{figure}[t]
\centering
\includegraphics[width=0.75\linewidth]{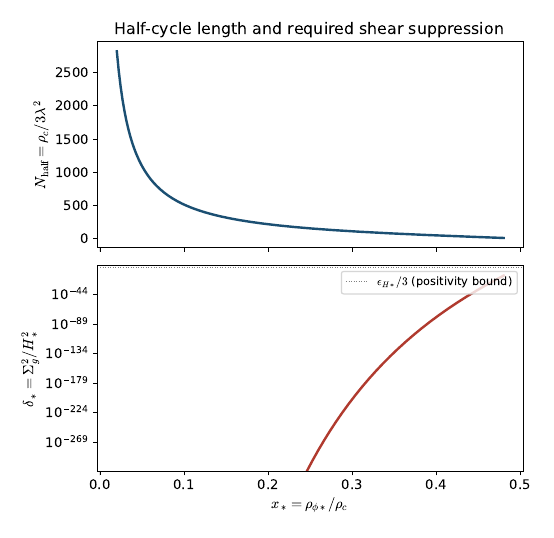}
\caption{Half-cycle length $N_{\rm half}=\rho_c/3\lambda^2$ (top) and the resulting
required residual shear fraction $\delta_*$ (bottom) as functions of the pivot
fraction $x_*$, for a bounce fixed at $\Omega_{\sigma B}=(\lambda^2/6)/10$. The dotted
line marks the theoretical positivity bound $\epsilon_{H*}/3$ of
Eq.~\eqref{eq:pivot-eqs}. The two points $x_*=0.1$ and $x_*=0.4$ discussed explicitly
in the text lie on this curve.}
\label{fig:tuning}
\end{figure}

As shown above, CMB data can be used to fix the parameters of the present cyclic-cosmology construction on the anisotropic 3-brane. We now carry out this normalization procedure explicitly. We fix the fiducial values $A_s=2.1\times10^{-9}$, $n_{s*}=0.965$, so that
Eq.~\eqref{eq:tilt} gives $\epsilon_{H*}=8.674\times10^{-3}$. Taking $x_*=0.1$ and, for the
moment, drop $\delta_*$ at leading order (it will emerge below as parametrically tiny, so this
is self-consistent). Equations~\eqref{eq:Hstar}--\eqref{eq:rhoc-fit} give
\begin{equation}
H_*^2 \simeq 1.80\times10^{-9}, \qquad
\lambda^2 \simeq 3.90\times10^{-11}, \qquad
\rho_c \simeq 5.99\times10^{-8}, \qquad
\rho_{\phi*}=x_*\rho_c\simeq5.99\times10^{-9}
\label{eq:worked-cmb}
\end{equation}
in reduced Planck units ($M_P=1$). For definiteness we record here the dimensions
of every quantity appearing in Eqs.~\eqref{eq:worked-cmb} and \eqref{eq:worked-x04},
since not all of them are scale-invariant. With $8\pi G_N=M_P^2=1$ and
$M_P=2.435\times10^{18}\,$GeV, the scalar field $\phi$ carries dimension $[M_P]$, so
that $\lambda=|\dot\phi|$ carries $[M_P^2]$ and $\lambda^2$ carries $[M_P^4]$; energy
densities $\rho_\phi$, $\rho_c$ and the brane tension $|\sigma|=\rho_c/2$ carry
$[M_P^4]$; $H_*^2$, $\Omega_\sigma$ and $\Omega_{\sigma B}$ carry $[M_P^2]$; and
$\Sigma_g^2=a_*^6\Omega_{\sigma*}$ carries $[M_P^2]$ for the dimensionless scale-factor
normalization adopted here. The remaining quantities --- $x_*$, $x_{\rm end}$,
$\epsilon_{H*}$, $n_{s*}$, $A_s$, $\delta_*$ and all $e$-fold numbers --- are
dimensionless. Restoring units, the benchmark of Eq.~\eqref{eq:worked-cmb} corresponds
to
\begin{equation}
H_*\simeq4.2\times10^{-5}M_P\simeq1.0\times10^{14}\,{\rm GeV},
\quad
\lambda\simeq6.2\times10^{-6}M_P^2,
\quad
\rho_c^{1/4}\simeq1.6\times10^{-2}M_P\simeq3.8\times10^{16}\,{\rm GeV},
\label{eq:worked-units}
\end{equation}
i.e.\ an inflationary Hubble scale and a brane tension scale
$|\sigma|^{1/4}=(\rho_c/2)^{1/4}\simeq3.2\times10^{16}\,$GeV both comfortably
sub-Planckian, as required for the effective four-dimensional description to be
trustworthy.

With these parameter values obtained from CMB-normalization the bounce location can be predicted, it does not remain a free input. Once $\lambda^2$ and $\rho_c$ are fixed, Eq.~\eqref{eq:rhophi-of-N} pins down how many $e$-folds
separate the pivot from the bounce: 
\begin{equation}
\Delta N_B \ \equiv\ N_*-N_B \ \simeq\ \frac{\rho_c(1-x_*)}{3\lambda^2}
\ \simeq\ \frac{5.39\times10^{-8}}{1.17\times10^{-10}} \ \simeq\ 461 .
\label{eq:worked-DeltaNB}
\end{equation}
This number is fixed by $A_s$, $n_{s*}$, and $x_*$ alone; it cannot be tuned independently.
In the above calculation we have assumed the weak-shear regime ($\rho_{\phi B}\simeq\rho_c$, cf.\
Eq.~\eqref{eq:rhoB-approx}). In our convention $x=\rho_\phi/\rho_c$, and consequently at the end of inflation when we $x=x_{\rm end}$, we have $\epsilon_H(x_{\rm end})=1$. If we neglect shear, ordinary inflationary theory gives (using Eq.~(\ref{eq:epsilonH-isotropic}))
$$\epsilon_H \simeq \frac{3\lambda^2}{2\rho_c}\frac{1-2x}{x(1-x)}\,,$$
where one can impose $\epsilon_H(x_{\rm end})=1$ and get the equation for $x_{\rm end}$:
\begin{eqnarray}
x_{\rm end}^2 - \left(1+ \frac{3\lambda^2}{\rho_c}\right)x_{\rm end} + \frac{3\lambda^2}{2\rho_c}=0\,.
\label{xende}
\end{eqnarray}
From the benchmark values of $\lambda^2$ and $\rho_c$ as given above we can safely neglect $3\lambda^2/\rho_c$ compared to unity, and the above equation becomes:
\begin{eqnarray}
x_{\rm end}^2 - x_{\rm end} + \frac{3\lambda^2}{2\rho_c}=0\,.
\label{xende1}
\end{eqnarray}
We can now calculate the number of $e$-folds from the pivot point to the end of inflation.  As the system evolves from the pivot point $\rho_\phi$ falls to
$\rho_{\phi,\rm end}\simeq x_{\rm end}\rho_c$ with $x_{\rm end}$ the small root of
of the above equation, giving $x_{\rm end}\simeq9.8\times10^{-4}$ and
\begin{equation}
\Delta N_{\rm end}\ \equiv\ N_{\rm end}-N_* \ \simeq\ \frac{\rho_{\phi*}-\rho_{\phi,\rm end}}{3\lambda^2}
\ \simeq\ 50.7 ,
\label{eq:worked-DeltaNend}
\end{equation}
where $N_{\rm end}$ is where $\epsilon_H=1$. That $\Delta N_{\rm end}$ comes out close to the
canonical $50$--$60$ e-fold window, without being tuned to do so, is a nontrivial consistency
check of this parameter point rather than an input. In our case we have
$$\Delta N_{\rm half}=N_T - N_B = \frac{x_B-x_T}{(3\lambda^2/\rho_c)}\,.$$
For $x_*=0.1$ and our benchmark values we get $\lambda^2/\rho_c \sim 6.51 \times 10^{-4}$ and for weak shear approximation at the turning points, we have $x_B \simeq 1$ and $x_T \simeq 0$. These numbers then yield $\Delta N_{\rm half} \simeq 512$.

The magnitude of $\Delta N_B\simeq461$ obtained in Eq.~\eqref{eq:worked-DeltaNB} deserves
comment, since at first sight such a large number of $e$-folds before the pivot appears
to leave no room for an ordinary inflationary era of the standard duration. In fact the
entire $e$-fold budget of the expanding branch is fixed analytically, and it is
instructive to display it. Dividing Eq.~\eqref{eq:rhoc-fit} by Eq.~\eqref{eq:lambda2} at
$\delta_*\to0$ gives the exact CMB-normalized ratio
\begin{equation}
\frac{\rho_c}{3\lambda^2} \;=\; \frac{1-2x_*}{2\,\epsilon_{H*}\,x_*(1-x_*)},
\label{eq:rhoc-over-lambda2}
\end{equation}
which depends on the observations only through $\epsilon_{H*}=(1-n_{s*})/(5-n_{s*})$.
Substituting Eq.~\eqref{eq:rhoc-over-lambda2} into Eq.~\eqref{eq:rhophi-of-N} yields
closed-form expressions for each segment of the cycle:
\begin{align}
\Delta N_{\rm sinf} &= \frac{\rho_c}{6\lambda^2}
= \frac{1-2x_*}{4\,\epsilon_{H*}\,x_*(1-x_*)} , \label{eq:efolds-sinf}\\[2pt]
\Delta N_{B} &= \frac{\rho_c(1-x_*)}{3\lambda^2}
= \frac{1-2x_*}{2\,\epsilon_{H*}\,x_*} , \label{eq:efolds-B}\\[2pt]
\Delta N_{\rm end} &\simeq \frac{\rho_{\phi*}}{3\lambda^2}
= \frac{1-2x_*}{2\,\epsilon_{H*}(1-x_*)} , \label{eq:efolds-end}
\end{align}
the last holding up to the small correction $x_{\rm end}\simeq3\lambda^2/2\rho_c\ll1$.
Three consequences follow.

First, the observationally relevant number is $\Delta N_{\rm end}$, the number of
$e$-folds between the pivot exit and the end of inflation, and
Eq.~\eqref{eq:efolds-end} shows that it is \emph{bounded above} by
$1/2\epsilon_{H*}\simeq57.6$ for any admissible $x_*$, and depends on $n_{s*}$ alone
through $\epsilon_{H*}$. It automatically lands inside the canonical $50$--$60$ window
provided $x_*\lesssim0.117$; this is not an input but a prediction of the CMB
normalization, and it is the sense in which the observable inflationary era of the model
is of ``simple'' standard duration.

Second, the large $\Delta N_B$ is \emph{not} an equally large ordinary-inflationary
era, and in particular it is not a superinflationary era either. Decomposing
Eqs.~\eqref{eq:efolds-sinf}--\eqref{eq:efolds-end} at $x_*=0.1$,
\begin{equation}
\underbrace{\Delta N_{\rm sinf}\simeq256}_{\epsilon_H<0,\ \ \rho_\phi>\rho_c/2}
\;+\;
\underbrace{204.8}_{0<\epsilon_H\ll1,\ \text{pre-pivot}}
\;=\;\Delta N_B\simeq461,
\qquad
\underbrace{\Delta N_{\rm end}\simeq50.7}_{\text{observable window}} ,
\label{eq:efold-budget}
\end{equation}
so that superinflation accounts for just over half of the pre-pivot interval and the
remainder is ordinary slow-roll inflation whose modes exited the horizon long before the
observable window and are today far outside it. Nothing observable is therefore affected
by the size of $\Delta N_B$; what it does affect is the shear amplification factor
$e^{6\Delta N_B}$, which is precisely the origin of the anisotropy problem quantified in
Sec.~\ref{subsec:strategy}.

Third, $\Delta N_B$ \emph{can} be controlled, but only at a price that
Eqs.~\eqref{eq:efolds-B}--\eqref{eq:efolds-end} make explicit. Since
$\Delta N_B/\Delta N_{\rm end}=(1-x_*)/x_*$, shortening the pre-pivot stage by pushing
$x_*$ towards $1/2$ necessarily shortens the observable stage in the same proportion:
at $x_*=0.4$ one finds $\Delta N_B\simeq28.8$ but only $\Delta N_{\rm end}\simeq19$,
which is too short to solve the horizon problem in the usual way. Within the
strict uniform-rate ansatz the two requirements --- a short pre-pivot interval and
$50$--$60$ observable $e$-folds --- are therefore mutually exclusive, and demanding
$\Delta N_{\rm end}\ge50$ forces $\Delta N_B\gtrsim377$. This tension is not an accident
of the parameter choice but a direct consequence of $d\rho_\phi/dN$ being a constant,
and it is one of the motivations for the generalized ansatz
$\dot\phi=-\lambda f(N)$ introduced in Sec.~\ref{subsec:extension}: since
$\Delta N=\int d\rho_\phi/(3\lambda^2f^2)$, a profile with $f^2>1$ in the
high-density segment between the bounce and the pivot compresses that segment while
leaving $f\simeq1$, and hence the CMB predictions, untouched in the observable window.
Taking $f^2\simeq10$ there would reduce $\Delta N_B$ from $461$ to $\simeq46$ and soften
the required shear suppression from $\delta_*\sim10^{-1205}$ to
$\delta_*\sim10^{-124}$ --- a dramatic improvement which nonetheless, as emphasised
throughout, does not by itself eliminate the tuning.

\subsubsection{Required shear amplitude and some other details}

The bounce inequality~\eqref{eq:bounce-condition-uniform}, evaluated at $\rho_{\phi B}\simeq\rho_c$,
caps the shear at the bounce at 
\begin{eqnarray}
\Omega_{\sigma B}<\lambda^2/6\simeq6.5\times10^{-12}\simeq1.08\times10^{-4}\rho_c\,.
\end{eqnarray}
This condition is does not depend upon how one normalizes the scale factor $a(t)$.
We choose the bounce comfortably inside this bound, $\Omega_{\sigma B}=10^{-5}\rho_c\simeq5.99\times10^{-13}$
(a margin of $\sim11$), safely satisfying~\eqref{eq:bounce-condition-uniform}. Since
$\Omega_\sigma(N)=\Sigma_g^2 e^{-6(N-N_*)}$ with $a_*=1$, and the bounce sits $\Delta N_B\simeq461$
$e$-folds before the pivot,
\begin{equation}
\Sigma_g^2 = \Omega_{\sigma B}\,e^{-6\Delta N_B} \ \simeq\ 6.0\times10^{-13}\times10^{-1201}
\ \simeq\ 6\times10^{-1214},
\qquad
\delta_* = \frac{\Sigma_g^2}{H_*^2} \ \simeq\ 3\times10^{-1205} .
\label{eq:worked-delta}
\end{equation}
This confirms self-consistently that $\delta_*\ll1$ was the correct assumption used in
Eq.~\eqref{eq:worked-cmb}. It also exhibits, in explicit numbers, the severity of the
shear-suppression problem flagged qualitatively previously: because
the CMB-normalized $\lambda^2/\rho_c$ is small, the self-consistent gap between bounce and pivot
is large ($\sim460$ e-folds), and over that many e-folds the $a^{-6}$ shear-dilution factor
amplifies any residual shear by $\sim10^{1201}$. Keeping the bounce weak-shear therefore forces
$\Sigma_g^2$ to be tuned to a value some $1200$ orders of magnitude below $H_*^2$. The only point to note here is that, the value of $\Sigma_g^2$ quoted above depends on the normalization of the scale factor $a(t)$. In some other way of normalization of the scale factor, one can make $\Sigma_g^2=1$.

The size of this required suppression is not fixed by the model; it is controlled by $\Delta N_B$, which shrinks as $x_*\to\tfrac12$. Repeating the same
steps at $x_*=0.4$ (same $A_s$, $n_{s*}$, and $10^{-5}\rho_c$ bounce margin) gives
\begin{equation}
H_*^2\simeq7.19\times10^{-9}, \quad
\lambda^2\simeq6.24\times10^{-10}, \quad
\rho_c\simeq8.99\times10^{-8}, \quad
\Delta N_B\simeq28.8, \quad
\delta_*\simeq1.3\times10^{-79} .
\label{eq:worked-x04}
\end{equation}
Here, as in Eq.~\eqref{eq:worked-cmb}, $H_*^2$ is in units of $M_P^2$, $\lambda^2$ and
$\rho_c$ in units of $M_P^4$, while $\Delta N_B$ and $\delta_*$ are dimensionless; in
physical units this point corresponds to $H_*\simeq2.1\times10^{14}\,$GeV and
$\rho_c^{1/4}\simeq4.2\times10^{16}\,$GeV. The accompanying $e$-fold budget, from
Eqs.~\eqref{eq:efolds-sinf}--\eqref{eq:efolds-end}, is
$\Delta N_{\rm sinf}\simeq24$, $\Delta N_{\rm half}\simeq48$ and
$\Delta N_{\rm end}\simeq19$, the last being the price paid for the shorter pre-pivot
interval, as discussed above.
Moving $x_*$ from $0.1$ to $0.4$ reduces the required tuning from $\sim10^{-1205}$ to
$\sim10^{-79}$ — enormously, but the tuning does not disappear: it remains many tens of orders
of magnitude below any value $\Sigma_g^2$ could plausibly take without a dedicated mechanism to
suppress it like the slow ekpyrotic contraction \cite{Ijjas} alternative to inflation. We conclude that the qualitative conclusion about the possibility of anistropy suppression
is robust to the choice of pivot point: some suppression mechanism for $\Sigma_g^2$, beyond
simply choosing $x_*$ close to $\tfrac12$, is required for this cyclic construction to be
observationally viable.

With $\Sigma_g^2$ this small, $\Omega_\sigma(N)\ll H^2(N)$ over the entire cycle except within a
narrow window of order a few e-folds on either side of $N_B$ and $N_T$ (where the exponential
$e^{-6(N-N_*)}$ compensates the smallness of $\Sigma_g^2$). Outside these two narrow windows the
background is numerically indistinguishable from the isotropic uniform-rate brane solution, so
$V(\phi)$ coincides there with the closed-form cosine potential of the isotropic
model. The periodicity condition~\eqref{eq:periodicity-condition} is therefore automatically
satisfied over the bulk of the cycle by inheritance from the isotropic solution; what remains is
to patch the two brief anisotropic transition regions near $N_B$ and $N_T$ so that $V(\phi)$ and
$V_{,\phi}$ match continuously onto the isotropic branch on both sides. We have not carried out
this patching explicitly; it is a short numerical boundary-value problem restricted to the
$O(1)$-e-fold neighborhoods of the two turning points, rather than an obstruction over the full
$\sim500$-e-fold cycle.
\subsection{A possible extension of the uniform rate model and possibilities of further refinements}
\label{subsec:extension}

In the minimal scalar-field realization discussed above, the scalar was assumed to obey the constant-rate ansatz
$\dot\phi=-\lambda={\rm constant}$, yielding the condition $d\rho_\phi/dN=-3\lambda^2$. Thus, in the constant-rate model, the scalar density decreases linearly with the number of e-folds,
\begin{equation}
    \rho_\phi(N)=\rho_{\phi *}-3\lambda^2(N-N_*).
\end{equation}
This feature is useful because it makes the background reconstruction simple, but it also implies that once the scalar reaches the end of inflation, it reaches the negative-density turnaround rather quickly. In particular, for the CMB-normalized branch with the choice $x_*=0.1$, it is found that:
\begin{equation}
\Delta N_{\rm end\to T}\equiv N_T-N_{\rm end}\simeq \frac12.
\end{equation}
Thus the original constant-rate model has essentially no room for a long post-inflationary stage. This motivates a mild generalization of the scalar ansatz.

Henceforth, we replace the constant-rate ansatz by the condition:
\begin{equation}
    \dot\phi=-\lambda f(N),
    \label{eq:variable-rate-ansatz}
\end{equation}
where $f(N)$ is a dimensionless profile. The constant-rate model is recovered for
\begin{equation}
    f(N)=1.
\end{equation}
For the generalized ansatz,
\begin{equation}
    \rho_\phi+p_\phi=\dot\phi^2=\lambda^2 f^2(N),
\end{equation}
and hence the scalar conservation equation becomes
\begin{equation}
\frac{d\rho_\phi}{dN}=-3\lambda^2 f^2(N).
\label{eq:variable-rate-rho-evolution}
\end{equation}
This is the central mechanism of our extension. If $f^2(N)$ is close to unity during the CMB-relevant inflationary regime, then the inflationary calculation remains essentially unchanged. If, however, $f^2(N)$ becomes much smaller after the end of inflation, then the scalar density decreases much more slowly and the interval between the end of inflation and the turnaround is elongated.

The background Friedmann equation is kept in the same form, and the geometric shear contribution also does not change in this new ansatz.
The scalar potential is reconstructed from $\rho_\phi=\frac12\dot\phi^2+V$,
which gives
\begin{equation}
V(N)=\rho_\phi(N)-\frac12\lambda^2 f^2(N).
\label{eq:potential-variable-rate}
\end{equation}
The scalar field itself is obtained parametrically. From Eq.~\eqref{eq:variable-rate-ansatz} and $dN=Hdt$,
\begin{equation}
    \frac{d\phi}{dN}=-\frac{\lambda f(N)}{H(N)},
\end{equation}
and hence
\begin{equation}
\phi(N)=\phi_0-\lambda\int^N\frac{f(\bar N)}{H(\bar N)}\,d\bar N.
\label{eq:phi-variable-rate}
\end{equation}
Equations~\eqref{eq:potential-variable-rate} and \eqref{eq:phi-variable-rate} provide a parametric reconstruction of the potential $V(\phi)$.

For the ansatz \eqref{eq:variable-rate-ansatz},
\begin{equation}
    \ddot\phi=-\lambda H f_{,N}.
\end{equation}
Therefore the Klein--Gordon equation requires
\begin{equation}
    -\lambda H f_{,N}-3\lambda H f+V_{,\phi}=0,
\end{equation}
or
\begin{equation}
V_{,\phi}=\lambda H\left(f_{,N}+3f\right).
\label{eq:KG-reconstruction-variable-rate}
\end{equation}
When $f=1$, this reduces to the constant-rate relation $V_{,\phi}=3\lambda H.$ Thus the generalized construction contains the constant-rate solution as a special case.

A useful smooth profile is:
\begin{equation}
f^2(N)=\eta+(1- \eta)\frac{1-\tanh\left[(N-N_s)/\Delta\right]}{2}
\label{eq:smooth-f-profile}
\end{equation}
with
\begin{equation}
    0<\eta\ll1\,,
\end{equation}
where $\eta$ is a new parameter of the theory.
Here $N_s$ denotes the transition point and $\Delta$ controls the smoothness of the transition. For $N\ll N_s$, one has
\begin{equation}
    f^2(N)\simeq1,
\end{equation}
so the CMB and ordinary inflationary epoch are effectively unchanged. For $N\gg N_s$, one has
\begin{equation}
    f^2(N)\simeq\eta,
\end{equation}
and the density evolution becomes
\begin{equation}
    \frac{d\rho_\phi}{dN}\simeq -3\lambda^2\eta.
\end{equation}
Therefore the same decrease in scalar density requires roughly $1/\eta$ more $e$-folds than in the constant-rate model.

This gives a simple estimate for the elongation. After inflation, suppose that $f^2\simeq\eta$ and that the turnaround occurs at a very small negative scalar density. Then
\begin{equation}
    \Delta N_{\rm end\to T}
    \simeq
    \frac{\rho_{\phi,\rm end}-\rho_{\phi T}}{3\lambda^2\eta}.
    \label{eq:Nend-to-T-general}
\end{equation}
In the shear-negligible end-of-inflation estimate, $\rho_{\phi,\rm end}\simeq \frac{3\lambda^2}{2}$,
while for the low energy density turnaround, $\rho_{\phi T}\simeq 0^-$.
Equation~\eqref{eq:Nend-to-T-general} then gives
\begin{equation}
\Delta N_{\rm end\to T}\simeq \frac{1}{2\eta}.
\label{eq:elongation-main-result}
\end{equation}
The original constant-rate model corresponds to $\eta=1$ and hence gives $\Delta N_{\rm end\to T}\simeq 1/2$. By contrast, if
$\eta=10^{-2}$, then $\Delta N_{\rm end\to T}\simeq 50$,
while $\eta=5\times 10^{-3}$ gives $\Delta N_{\rm end\to T}\simeq 100$.
Thus a modest suppression of the scalar velocity after inflation greatly elongates the post-inflationary branch.

The CMB normalization is not spoiled if the transition in $f(N)$ occurs after the CMB-relevant inflationary window. In the generalized model, the curvature perturbation amplitude in the effective shear-negligible single-field regime is
\begin{equation}
    \mathcal P_{\mathcal R}
    \simeq
    \frac{H_*^4}{4\pi^2\lambda^2 f_*^2}.
    \label{eq:PR-variable-rate}
\end{equation}
If
\begin{equation}
    f_*\simeq1,
\end{equation}
then Eq.~\eqref{eq:PR-variable-rate} reduces to the constant-rate result
\begin{equation}
    \mathcal P_{\mathcal R}
    \simeq
    \frac{H_*^4}{4\pi^2\lambda^2}.
\end{equation}
The inflationary regime is left essentially unaffected provided the parameter $\eta$ is chosen appropriately.
The turnaround condition remains $H^2(N_T)=0$, as before.
Since it gives $\Omega_\sigma\propto e^{-6N}$, elongating the post-inflationary interval makes $\Omega_{\sigma T}$ even smaller.
The turnaround therefore occurs at a very small negative scalar density,
\begin{equation}
    \rho_{\phi T}\simeq0^-.
\end{equation}
This confirms that the post-inflationary phase can be elongated without violating the bounce/turnaround results derived above.

For an exactly cyclic completion, one must also impose a recurrence condition. Over the expanding half-cycle,
\begin{equation}
    \Delta\rho_\phi^{\rm exp}
    =
    -3\lambda^2\int_{N_B}^{N_T} f_{\rm exp}^2(N)\,dN.
\end{equation}
Over the contracting half-cycle, written over the same positive interval,
\begin{equation}
    \Delta\rho_\phi^{\rm con}
    =
    +3\lambda^2\int_{N_B}^{N_T} f_{\rm con}^2(N)\,dN.
\end{equation}
Exact recurrence of the scalar density requires
\begin{equation}
    \int_{N_B}^{N_T} f_{\rm exp}^2(N)\,dN
    =
    \int_{N_B}^{N_T} f_{\rm con}^2(N)\,dN.
    \label{eq:recurrence-condition-f}
\end{equation}
A simple sufficient condition is to choose the two profiles to be the same after the appropriate reflection between the expanding and contracting branches. In addition, the reconstructed potential should satisfy endpoint matching,
\begin{equation}
    V_{\rm final}=V_{\rm initial},
    \qquad
    V_{,\phi}^{\rm final}=V_{,\phi}^{\rm initial},
\end{equation}
and, for a globally periodic scalar realization,
\begin{equation}
    V(\phi+\Delta\phi_{\rm cyc})=V(\phi).
\end{equation}
These matching conditions are not automatic consequences of the local elongation mechanism; they must be imposed or checked in the reconstructed cyclic potential.

Finally, this generalized scalar branch should not by itself be interpreted as a complete hot-big-bang history. The elongated segment is a scalar-field drift phase, not automatically a radiation- or matter-dominated era. A realistic completion would require reheating and ordinary matter/radiation sectors. In a cyclic setting, reheating must be incomplete: a residual scalar component must survive, remain sufficiently hidden during the subsequent cosmological evolution, and later enter a negative-potential region so as to trigger the large-radius turnaround.

\section{The isotropic limit: exact bounce, turnaround, and periodicity at $\Sigma_g=0$}
\label{subsec:isotropic-limit}

It is instructive to switch off the geometric shear entirely, $\Sigma_g=0$, $\Omega_\sigma\equiv0$.
This is not the regime of physical interest for the anisotropic construction, but it is exactly
solvable, and it isolates precisely which features are caused by the shear versus which are intrinsic to the uniform-rate scalar on the minimal
timelike brane.

With $\Omega_\sigma=0$, Eq.~\eqref{eq:master-friedmann} becomes
\begin{equation}
H^2 = \frac{\rho_\phi}{3}\left(1-\frac{\rho_\phi}{\rho_c}\right).
\label{eq:isotropic-F}
\end{equation}
This is manifestly non-negative only for $0\le\rho_\phi\le\rho_c$, and vanishes exactly at the
two endpoints
\begin{equation}
\rho_\phi = 0 \qquad \text{and} \qquad \rho_\phi = \rho_c .
\label{eq:isotropic-turning}
\end{equation}
These are precisely the two limiting points of the general no-turning-point interval
$0<\rho_*<\rho_c$ established earlier: switching off the shear does
not create a qualitatively new turning-point structure, it shrinks the strict inequalities
$\rho_*<0$ or $\rho_*>\rho_c$ of Eq.~\eqref{eq:two-branches} down onto their boundary values.

Setting $\Omega_{\sigma*}=0$ in the general formula Eq.~\eqref{eq:Hdot-uniform} gives
\begin{equation}
\dot H_* = -\frac{\lambda^2}{2}\left(1-\frac{2\rho_{\phi*}}{\rho_c}\right).
\label{eq:Hdot-isotropic}
\end{equation}
At the two turning points this evaluates to
\begin{equation}
\dot H\big|_{\rho_\phi=\rho_c} = +\frac{\lambda^2}{2}, \qquad
\dot H\big|_{\rho_\phi=0} = -\frac{\lambda^2}{2},
\label{eq:Hdot-isotropic-values}
\end{equation}
equal in magnitude and opposite in sign. Comparing with the bounce condition
Eq.~\eqref{eq:bounce-condition-uniform}, whose right-hand side is $6\Omega_{\sigma B}$: with
$\Omega_{\sigma B}=0$ this inequality is satisfied automatically, with no competition to
arrange. Thus $\rho_\phi=\rho_c$ is \emph{always} a bounce and $\rho_\phi=0$ is \emph{always}
a turnaround for the isotropic uniform-rate scalar on this brane --- no tuning of any parameter
is required, in sharp contrast to the anisotropic case.

Since $d\rho_\phi/dN=-3\lambda^2$ is constant (Eq.~\eqref{eq:rhophi-of-N}) and $\rho_\phi$ sweeps
the full interval $[0,\rho_c]$ exactly once per half-cycle,
\begin{equation}
\Delta N_{\rm half} = \frac{\rho_c}{3\lambda^2}
\label{eq:isotropic-Nhalf}
\end{equation}
holds \emph{exactly} (not merely at leading order in weak shear, as in
the anisotropic case). Because $\Omega_\sigma=0$ identically, the field equation can be integrated in closed form.
Writing $u\equiv\rho_\phi=\lambda^2/2+V$, Eq.~\eqref{eq:KG-uniform} together with
Eq.~\eqref{eq:isotropic-F} gives $(V_{,\phi})^2=9\lambda^2H^2=3\lambda^2u(1-u/\rho_c)$, i.e.
\begin{equation}
\left(\frac{du}{d\phi}\right)^2 = k^2\,u(\rho_c-u), \qquad k\equiv\sqrt{\frac{3\lambda^2}{\rho_c}} .
\label{eq:ODE-u}
\end{equation}
This is solved exactly by
\begin{equation}
\rho_\phi(\phi) = \frac{\rho_c}{2}\Big[1-\cos\big(k\phi+c_0\big)\Big],
\label{eq:trig-solution}
\end{equation}
with $c_0$ an integration constant, as is verified directly by differentiating
Eq.~\eqref{eq:trig-solution} and substituting into Eq.~\eqref{eq:ODE-u}. The reconstructed
potential is therefore
\begin{equation}
\ V(\phi) = \frac{\rho_c}{2}\Big[1-\cos\big(k\phi+c_0\big)\Big] - \frac{\lambda^2}{2}\,,
\qquad k=\sqrt{\frac{3\lambda^2}{\rho_c}} .
\label{eq:Vphi-isotropic}
\end{equation}
Equation~\eqref{eq:Vphi-isotropic} is manifestly periodic in $\phi$ with period
\begin{equation}
\Delta\phi_{\rm cyc} = \frac{2\pi}{k} = 2\pi\sqrt{\frac{\rho_c}{3\lambda^2}} ,
\label{eq:period-phi}
\end{equation}
and $V_{,\phi}=0$ exactly at $\rho_\phi=0,\rho_c$, consistent with Eq.~\eqref{eq:KG-uniform} at
the turning points. This is the exact realization of the periodicity condition
Eq.~\eqref{eq:periodicity-condition} left open previously: because $\phi(t)$
is strictly monotonic while $V(\phi)$ is genuinely periodic, $\rho_\phi(N)$ can cycle
indefinitely between $0$ and $\rho_c$ even though the field itself never reverses direction.

The isotropic relation \eqref{eq:isotropic-F} deserves a comment, because it is not
peculiar to the braneworld. It is functionally identical to the effective Friedmann
equation of loop quantum cosmology (LQC), $H^2=(\rho/3)(1-\rho/\rho_c^{\rm LQC})$, with
the brane critical density $\rho_c=2|\sigma|$ playing the role of the LQC critical
density $\rho_c^{\rm LQC}\simeq0.41\rho_{\rm Pl}$. This coincidence is the well-known
duality between loop cosmological dynamics and Randall--Sundrum-type braneworlds
\cite{SinghDuality}: the quadratic correction that a timelike extra dimension supplies
geometrically ($\epsilon=-1$) is the same correction that holonomy effects supply
quantum-mechanically, even though the two have entirely different origins --- one
classical and five-dimensional, the other four-dimensional and quantum. It follows that
any result derived from Eq.~\eqref{eq:isotropic-F} alone must be common to both
frameworks, and indeed the exactly solvable isotropic uniform-rate model has been
analyzed independently in the LQC setting by Paul and Saini \cite{PaulSaini}, who obtain
a cyclic universe with identical cycles separated by bounces at the critical density and
a reconstructed potential of the polymerized/hybrid-natural-inflation type. Our
Eqs.~\eqref{eq:isotropic-Nhalf}--\eqref{eq:period-phi} are the braneworld counterparts
of those results, and we acknowledge the agreement explicitly: within the isotropic
limit, the periodic cosine potential, the exact half-cycle length $\rho_c/3\lambda^2$
and the tuning-free alternation of bounce and turnaround are shared features, obtained
here via the $\epsilon=-1$ brane and there via holonomy corrections.

It is therefore important to state precisely where the new content of the present
work lies, since none of it is contained in the isotropic limit. (i) The entire analysis
of Sec.~\ref{sec:bounce-turnaround-cyclic} is \emph{anisotropic}: the Bianchi-I shear
term $\Omega_\sigma\propto a^{-6}$ is absent from Eq.~\eqref{eq:isotropic-F} and from
the isotropic LQC treatment, and it is the competition between this term and the
negative brane correction that produces every result of this paper. (ii) Once shear is
present, the symmetry between the two turning points is broken: we prove, for
\emph{any} fluid obeying the NEC, that the turnaround at $\rho<0$ remains unconditional
while the bounce becomes \emph{conditional} on the inequality
\eqref{eq:bounce-condition-general}. In the isotropic case both are automatic, so this
classification has no isotropic analogue. (iii) The bounce is displaced above the
critical density by a calculable amount, $\rho_{\phi B}\simeq\rho_c+3\Omega_{\sigma B}$
(Eq.~\eqref{eq:rhoB-approx}), whereas in LQC and in the isotropic brane limit the bounce
sits exactly at $\rho_c$. (iv) The turnaround is displaced to a small negative density,
$\rho_{\phi T}\simeq-3\Omega_{\sigma T}$, instead of exactly zero. (v) We derive the
finite-cyclic-branch matching condition \eqref{eq:matching-condition}, the
superinflationary duration \eqref{eq:superinf-duration} and its brane-tension scaling
\eqref{eq:sigma-scaling}, and the closed-form $e$-fold budget
\eqref{eq:efolds-sinf}--\eqref{eq:efolds-end}. (vi) Most importantly, we quantify the
anisotropy problem: the CMB-normalized requirement
$\delta_*\sim10^{-1205}$--$10^{-79}$ has no counterpart in an isotropic treatment, where
$\Sigma_g=0$ by assumption and the model is tuning-free. The isotropic section is
included here only as an exactly solvable consistency benchmark and as the limit in
which contact with \cite{PaulSaini} can be made; the physics we wish to report is the
obstruction that appears as soon as that limit is relaxed.

Equation~\eqref{eq:Vphi-isotropic} is the $\epsilon=-1$ counterpart of the hyperbolic-cosine
potential found for uniform-rate inflation on the (spacelike, $\epsilon=+1$) Randall--Sundrum
brane. Formally continuing $\rho_c\to-2\Lambda$ turns
$k=\sqrt{3\lambda^2/\rho_c}\to i\sqrt{3\lambda^2/2\Lambda}$ and $\cos\to\cosh$, reproducing
$V(\phi)=\Lambda\cosh\!\big(\sqrt{3\lambda^2/2\Lambda}\,\phi\big)-\Lambda-\lambda^2/2$. The
bounded, oscillatory potential of Eq.~\eqref{eq:Vphi-isotropic} versus the unbounded, monotonic
$\cosh$ potential of the spacelike case is the direct imprint, at the level of the reconstructed
potential, of the sign flip $\epsilon=-1$ that permits a bounce in the first place.

This isotropic point is not merely a mathematical curiosity: comparing
Eq.~\eqref{eq:isotropic-Nhalf} to the weak-shear estimate of Eq.~\eqref{eq:worked-DeltaNB}, and
noting that the bounce inequality \eqref{eq:bounce-condition-uniform} holds with strict positive
margin proportional to $\Omega_{\sigma B}$ once shear is switched back on, shows that \emph{all}
of the fine-tuning quantified in the earlier section and Fig.~\ref{fig:tuning} is
sourced entirely by $\Sigma_g\neq0$. The $\Sigma_g=0$ point is the exactly-solvable,
tuning-free boundary of the parameter space; it is also, however, the point at which the model
contains no actual anisotropy, so it should be read as a consistency benchmark for the
construction of Sec.~\ref{sec:bounce-turnaround-cyclic} rather than as a
substitute for it.

\section{Conclusion and Discussion}

In this work we have constructed and analyzed a cyclic cosmological
scenario driven by a uniform-rate scalar field on an anisotropic Bianchi-I
brane embedded in a bulk with a timelike extra dimension, within the
Shtanov--Sahni braneworld framework. We summarize our main results and
discuss their physical significance below.

\subsection{Summary of results}

Starting from the general anisotropic Shtanov--Sahni brane Friedmann
equation, Eq.~\eqref{eq:SS-general}, we specialized to the flat,
dark-radiation-free, $\Lambda_{\rm eff}=0$ branch relevant for a timelike
extra dimension, obtaining the effective equation, Eq.~\eqref{eq:master-friedmann},
in which the negative quadratic density correction characteristic of a
timelike extra dimension competes against a positive, purely geometric
shear term $\Omega_\sigma(a)\propto a^{-6}$.

Imposing the uniform-rate condition $\dot\phi=-\lambda$ on the scalar
field, we showed that the energy density obeys a strictly linear
relation in the number of e-folds, Eq.~\eqref{eq:rhophi-of-N}, a result that
follows solely from energy conservation and is therefore independent of
the specific form of the Friedmann equation surviving unchanged from
the isotropic uniform-rate brane model into the present anisotropic
setting.

We then derived a completely general classification of turning points
for any fluid satisfying the null energy condition. Two qualitatively
different behaviors emerge: (i) turnarounds occurring at negative energy
density are \emph{unconditional}, requiring no fine-tuning of the shear
amplitude, Eq.~\eqref{eq:HdotT-negative}; (ii) bounces occurring at
$\rho>\rho_c$ are \emph{conditional}, requiring the high-energy brane
correction to dominate the decelerating shear term, Eq.~\eqref{eq:bounce-condition-general}.
This asymmetry is a direct and physically transparent consequence of the
opposite signs with which $\rho$ and $\Omega_\sigma$ enter the
Hamiltonian constraint away from and above $\rho_c$.

Specializing these general results to the uniform-rate scalar, we
obtained explicit bounce and turnaround conditions,
Eqs.~\eqref{eq:bounce-condition-uniform}--\eqref{eq:rhoT-approx}, showing that both the excess bounce
density above $\rho_c$ and the turnaround density near zero are
controlled, at leading order, by the local shear amplitude. Matching a
bounce to an adjacent turnaround yields the finite-cyclic-branch
condition, Eq.~\eqref{eq:matching-condition}, which becomes an explicit relation
between $\Omega_{\sigma B}$, $\Omega_{\sigma T}$, $\rho_c$, and the
half-cycle length $\Delta N$ in the weak-shear regime. Numerically
integrating $F(N)=H^2(N)$ confirms the expected sign structure at both
roots (Fig.~\ref{fig:turning-points}), validating the analytic classification.

On the expansion branch we identified two distinct accelerating phases:
a post-bounce superinflationary phase with $\epsilon_H<0$ and an
ordinary inflationary phase, $0<\epsilon_H<1$, that emerges once shear
has diluted and which reduces smoothly to the isotropic uniform-rate
result, Eq.~\eqref{eq:epsilonH-isotropic}. This is the regime in which the
single-field curvature power spectrum, Eqs.~\eqref{eq:PR-isotropic}--\eqref{eq:tilt},
and the associated tilt and running, Eqs.~\eqref{eq:running}--\eqref{eq:pivot-eqs},
are computed.

The duration of the superinflationary stage was obtained in closed form,
Eq.~\eqref{eq:superinf-duration}: it is terminated not by the shear, which
decays exponentially within a few $e$-folds of the bounce, but by the linear
decrease of $\rho_\phi$ through $\rho_c/2$, and it therefore occupies
$\Delta N_{\rm sinf}\simeq\rho_c/6\lambda^2$, exactly half the expanding
branch at leading order in weak shear. Written in terms of the brane tension
this is $\Delta N_{\rm sinf}\simeq|\sigma|/3\lambda^2$,
Eq.~\eqref{eq:sigma-scaling}, so that the high-energy stage lengthens linearly
with $|\sigma|$ while the maximum shear tolerated at the bounce,
$\Omega_{\sigma B}<\lambda^2/6$, is independent of it --- a larger brane
tension consequently worsens, rather than relieves, the anisotropy problem.
We also showed that once $A_s$ and $n_{s*}$ are fixed, the whole $e$-fold
budget of the cycle follows analytically from
Eqs.~\eqref{eq:efolds-sinf}--\eqref{eq:efolds-end}: the observable interval
$\Delta N_{\rm end}$ is bounded above by $1/2\epsilon_{H*}\simeq57.6$ and falls
naturally in the canonical $50$--$60$ window for $x_*\lesssim0.117$, while the
much larger $\Delta N_B$ consists of a superinflationary half and a pre-pivot
slow-roll half whose modes are unobservable today. Shortening $\Delta N_B$
within the strict uniform-rate ansatz shortens $\Delta N_{\rm end}$ in the same
proportion; escaping this trade-off requires the generalized profile
$\dot\phi=-\lambda f(N)$ of Sec.~\ref{subsec:extension}, with $f^2>1$ confined
to the high-density segment.

A central and, in our view, the most physically consequential result of
this paper concerns the fate of the anisotropy across a cycle. Because
$\Omega_\sigma\propto a^{-6}$, shear that dilutes during expansion is
reamplified by exactly the same factor during contraction,
Eq.~\eqref{eq:cyclic-shear-bound}, so that the shear inherited at the following
bounce must still satisfy the bounce inequality there,
Eq.~\eqref{eq:bounce-condition-uniform}. Working out an explicit CMB-normalized parameter
point ($A_s=2.1\times10^{-9}$, $n_{s*}=0.965$) at $x_*=0.1$, we found
that consistency between the pivot-scale observables and a long
($\sim500$ e-fold) half-cycle forces the dimensionless shear fraction at
the pivot to satisfy $\delta_*\equiv\Omega_{\sigma*}/H_*^2\sim
10^{-1205}$ --- an extraordinarily severe suppression, driven by the
$a^{-6}$ amplification acting over $\Delta N_B\simeq461$ e-folds between
bounce and pivot. Repeating the calculation at $x_*=0.4$ reduces this to
$\delta_*\sim10^{-79}$ (Sec.~\ref{subsec:strategy}, Fig.~\ref{fig:tuning}):
the required tuning is dramatically alleviated as $x_*\to1/2$, but it
never disappears entirely, and remains many tens of orders of magnitude
beyond what could plausibly arise without a dedicated suppression
mechanism.

Finally, we noted that reconstructing a strictly periodic potential
$V(\phi)$ requires not only that $\Delta V_{\rm cyc}=0$ around a cycle,
Eq.~\eqref{eq:DeltaVcyc}, but also the nontrivial matching condition,
Eq.~\eqref{eq:periodicity-condition}, of $V$ and $V_{,\phi}$ at the two field values
associated with a common value of $N$ on the expanding and contracting
half-cycles. In the phenomenologically relevant weak-shear regime this
condition is automatically satisfied over the bulk of the cycle, since
the background there is numerically indistinguishable from the isotropic
uniform-rate solution; what remains open is an explicit numerical
patching of $V(\phi)$ within narrow, $O(1)$-e-fold neighborhoods of the
bounce and turnaround, where the anisotropic corrections are not
negligible.

\subsection{Discussion}

The results above should be read as establishing both the internal
consistency and the principal phenomenological obstruction of
timelike-extra-dimension braneworld cyclicity. On the positive side, the
mechanism is geometrically transparent: the same negative quadratic
density correction that resolves the initial singularity also, generically
and without any additional matter content, produces a turnaround at
negative energy density with no tuning required. The uniform-rate
condition on $\phi$ is likewise minimal, requiring no engineered
potential; $V(\phi)$ is instead reconstructed self-consistently from the
background dynamics.

The obstruction is the well-known ``anisotropy problem'' of bouncing and
cyclic cosmologies, here rendered in closed analytic form,
Eq.~\eqref{eq:cyclic-shear-bound}, and quantified numerically for the first time in
this braneworld setting. Because $\Omega_\sigma$ scales as $a^{-6}$ while
$H^2$ scales, at low energies, as $a^{-3(1+w)}$ for ordinary matter, any
residual shear is parametrically enhanced relative to the background
during contraction; sustaining many hundreds of e-folds of low-curvature
evolution per half-cycle as demanded by consistency with the observed
scalar amplitude and tilt amplifies this effect to the point of
requiring $\Sigma_g^2$ to be tuned to a degree that has no natural
justification within the model as presently formulated. This is not a
flaw specific to the timelike-extra-dimension construction; analogous
tunings afflict essentially all bouncing and cyclic models with
non-vanishing anisotropic or curvature degrees of freedom. What is new
here is the explicit demonstration that the required suppression is a
steep, calculable function of the pivot density fraction $x_*$
(Fig.~\ref{fig:tuning}), which opens a concrete direction for future
work: rather than treating $\Sigma_g^2$ as a free initial condition, one
should seek a dynamical isotropization mechanism -- for example
brane-localized dissipative or bulk-viscous effects, a mild violation of
the closure assumption $C=0$ via a small dark-radiation term, or
higher-curvature (Gauss--Bonnet) corrections already invoked to stabilize
the timelike-extra-dimension construction against tachyonic Kaluza--Klein
modes that actively drives $\Omega_\sigma$ below the bounce threshold
independently of initial conditions. One can also take resort to slow ekpyrotic contraction mechanism of dynamical shear isotropization, abandoning the inflationary paradigm altogether and replacing it by the compelling ekpyrosis alternative within our braneworld framework.

It should be stressed, finally, that none of this obstruction is visible in the
isotropic limit. There the model is exactly solvable and completely tuning-free, and,
through the duality between the quadratic brane correction and holonomy corrections
\cite{SinghDuality}, it reproduces the uniform-rate cyclic solution and periodic
reconstructed potential found in loop quantum cosmology \cite{PaulSaini}. The
distinctive content of the present analysis is that the shear term, which is absent by
construction in that limit, converts an unconditional bounce into a conditional one,
displaces both turning points away from $\rho_\phi=\rho_c$ and $\rho_\phi=0$ by amounts
proportional to the local shear, and imposes the quantitative suppression requirement on
$\Sigma_g^2$ derived above.

A second direction concerns completing the periodicity analysis: an
explicit boundary-value patching of $V(\phi)$ and $H(\phi)$ across the
narrow anisotropic transition windows at $N_B$ and $N_T$ would confirm,
rather than infer by continuity, that the reconstructed potential is a
smooth, single-valued periodic function, and would additionally allow the
imprint of the brief superinflationary phase on the low-multipole CMB
spectrum to be assessed quantitatively. We leave both the search for a
natural shear-suppression mechanism and the explicit periodicity patching
for future work.

\section*{Acknowledgement}

RS dedicates this work to the memory of his beloved grandmother Kana Dutta who passed away on December 30, 2025. RS thanks the Council for Scientific and Industrial Research (CSIR), Government of India for financial support through the Research Associateship scheme.

\bibliography{Refs_corrected}

\end{document}